\documentclass{aa}

\usepackage{graphicx}
\usepackage{amsmath}
\usepackage{amssymb}
\usepackage{xspace}
\usepackage{commath}
\usepackage{times}
\usepackage{bm} 
\usepackage{balance}
\usepackage{hyperref}
\usepackage{mathtools}
\usepackage{stmaryrd}
\usepackage{tabulary}
\usepackage{xcolor}
\usepackage{txfonts}
\usepackage{orcidlink}

\usepackage{afterpage}
\usepackage{multicol}
\newsavebox{\shortpagebox}

\makeatletter
\newcommand{\shortpage}[1]
{\par
  \setbox\shortpagebox=\vbox{\strut #1\par}%
  \afterpage{\onecolumn
    \begin{multicols}{2}
    \unvbox\AP@partial
    \end{multicols}}%
  \unvbox\shortpagebox
\par}
\makeatother

\hypersetup{dvips, colorlinks=true, linkcolor=blue, citecolor=blue, filecolor=blue, urlcolor=blue}

\usepackage{xspace}
\usepackage{amsfonts,textcomp}

\newcommand{\ie}{\emph{i.e.}\@ifnextchar.{\!\@gobble}{}}
\newcommand{\eg}{\emph{e.g.}\@ifnextchar.{\!\@gobble}{}}
\newcommand{\etc}{etc\@ifnextchar.{}{.\@}}

\newcommand{\mg}{\left<}
\newcommand{\mdd}{\right>}

\newcommand{\Dirac}{\delta_{\textsc{d}}}
\newcommand{\tHyp}{_0{\tilde F}_1}

\newcommand{\vr}{{\bm r}}
\newcommand{\vk}{{\bm k}}
\newcommand{\vx}{{\bm x}}
\newcommand{\eps}{{\epsilon}}
\newcommand{\rhob}{\overline{\rho}}
\newcommand{\xib}{\overline{\xi}}
\newcommand{\mW}{\mathcal{C}}
\newcommand{\mR}{\mathcal{R}}
\newcommand{\hrho}{\hat{\rho}}
\newcommand{\hlambda}{\hat{\lambda}}
\newcommand{\lambdaR}{{\lambda^R}}
\newcommand{\ellS}{{\ell_\mathcal{S}}}
\newcommand{\flights}{Rayleigh-Levy flights\ }

\newcommand{\dd}{{\rm d}}
\newcommand{\ii}{{\rm i}}
\newcommand{\MTM}{{\rm MTM}}
\newcommand{\trees}{{\rm trees}}
\newcommand{\lines}{{\rm lines}}
\newcommand{\vertices}{{\rm vertices}}
\newcommand{\Gaussian}{{\rm Gaussian}}
\newcommand{\mixchi}{{\chi}}

\begin{document}


\title{The statistics of Rayleigh-Levy flight extrema}
\titlerunning{The statistics of Rayleigh-Levy flight extrema}
 \authorrunning{Francis Bernardeau \& Christophe Pichon}
\author{Francis Bernardeau\inst{1} 
\and Christophe Pichon\inst{1,2,3}\orcidlink{0000-0003-0695-6735}}
 \institute{
 Université Paris-Saclay, CNRS, CEA, Institut de physique théorique, 91191, Gif-sur-Yvette, France.
 \and
CNRS \& Sorbonne Universit\'e, Institut d'Astrophysique de Paris, 98 bis Boulevard Arago, 75014 Paris, France
 \and
Korea Institute for Advanced Studies (KIAS), 85 Hoegi-ro, Dongdaemun-gu, Seoul, 02455, Republic of Korea
}

\abstract{
Rayleigh-Levy flights have played a significant role in cosmology as simplified models for understanding how matter distributes itself under gravitational influence. These models also exhibit numerous remarkable properties that enable the prediction of a wide range of characteristics. Here, we derive the one and two point statistics of extreme points within Rayleigh-Levy flights spanning one to three dimensions, stemming directly from fundamental principles. In the context of the mean field limit, we provide straightforward closed-form expressions for Euler counts and their correlations, particularly in relation to their clustering behaviour over long distances. Additionally, quadratures allow for the computation of extreme value number densities. A comparison between theoretical predictions in 1D and Monte Carlo measurements shows  remarkable agreement. Given the widespread use of Rayleigh-Levy processes, these comprehensive findings offer significant promise not only in astrophysics but also in broader applications beyond the field.
}
\keywords{
Cosmology -- Clustering  -- large scale structures  -- Method: Analytical, numerical.
}

\maketitle

\section{Introduction}
\label{sec:introduction}

The geometry and structure of cosmic fields form a complex physical system that develops from homogeneity through the interplay of expansion and long-range forces. Its statistical properties should emulate those of such a class of systems. From the perspective of observational cosmology, its evolution mirrors both the history of the universe's expansion rate and the dynamic growth of embedded substructures \citep[see for instance][for an account of the emergence of structure due to gravitational instability] {1986ApJ...304...15B,Review2002}. 
While the small scale -- galaxy size -- structures are determined by the interaction of gravitational effects and complex baryonic physics, the large-scale structure is shaped solely by the development of the gravitational instabilities.

The most standard approach to describe the outcome of this evolution is to consider the matter correlation functions, or equivalently in Fourier space, the spectra \citep[e.g.][]{Peebles+1975,Fry1985,Bernardeau1994,Scoccimarro+1998,Cappi+2015}.
Those quantities capture however only partially the outcome of the gravitational processes. In particular gravitational instabilities leads to the formation of large scale structures, such as self gravitating massive haloes embedded in an intricate cosmic web made of voids, walls and filaments that are woven together under the influence of gravity \citep[][]{bondetal1996}.
Inspired by such emerging features, a related alternative to quantify the properties of the field is to explore 
the topology of the excursion of the cosmic web. It can be done 
both in real space \citep[e.g.][]{1994ApJ...434L..43M,Gay2012} and redshift space \citep[e.g.][]{1996ApJ...457...13M,2013MNRAS.435..531C}, using  tools such as the void probability function \citep[VPD, e.g.][]{White1979,sheth_hierarchy_2004}, the  Euler-Poincar\'e characteristic \citep[e.g.][]{GottMellotDickinson1986,ParkGott1991,appleby+18}, or more generally Minkowski functionals \citep[e.g.][]{MeckeBuchertWagner1994,SchmalzingBuchert1997},  persistent homology  \citep[e.g.][]{,Sousbie2011-2,Pranav+2017}, and Betti numbers \citep[e.g.][]{Park+2013,Feldbrugge+2019}.
Given the duality established by Morse-Smale theory \citep{Forman2002} between the geometry or topology of excursion,   and the loci of null gradients  of the underlying density field, a
summary statistics  is given by the point process of these critical points \citep[e.g.][]{1986ApJ...304...15B,bondetal1996,Gay2012,cadiou2020}. 
For instance,  there has recently been some interest in also using the clustering of such  points as cosmological probes  \citep{2020arXiv201214404B,shim+21}, e.g.,  extracted from  Lyman-$\alpha$ tomography \citep{Kraljic22}.  Unfortunately, the theory for capturing their statistics has been limited to the quasi Gaussian limit \citep[e.g.][]{2009PhRvD..80h1301P}, or slightly beyond \citep{2015MNRAS.449L.105B}, while relying on the large deviation principle. This puts limitation on its realm of application to larger  scales only (or involves relying on calibration over N-body simulations).

A notable  cosmologically  relevant counterexample  is provided by \flights\!\!\footnote{See \cite{Anomalous2008} for a fairly recent review.} \citep{1975CRASM.280.1551M,1980lssu.book.....P,1996ApJ...470..131S,2013ApJ...778...35Z,2019PAIJ....3...82U}, which are simply defined as a Markov chain point process whose jump probability depends on some power law of the length of the jump alone.
In 3D, it  acts as a coarse proxy to describing the motion of dark halos \citep{2005PhRvD..71f3001S,2015JPlPh..81a3208T}, but its definition can be extended to arbitrary dimensions. 
 \flights were first introduced in astrophysics by \cite{Holtsmark1919}   in the context of fluctuations in gravitational systems \citep{Litovchenko2021}.
They have some connection first passage theory \citep{10.1088/1742-5468/ab4988}, underlying Press Schechter theory for halo formation   \citep{10.1086/152650},  and subsequent mass accretion \citep{2018MNRAS.476.4877M}. 
\flights have also attracted lots of attention beyond cosmology, \citep[from anomalous cosmic rays diffusion][to ISM scintillation]{1999NuPhS..75..191W,2003ApJ...584..791B} or indeed beyond astronomy \citep[from turbulence,][to earthquakes]{1987PhRvL..58.1100S,2000GeoRL..27.1965S}, and even biology \citep[e.g.][]{10.1088/2399-6528/aad498} or risk management \citep[e.g.][]{Bouchaud2003},
as it captures anomalous diffusion \citep[see, e.g.][and references therein]{1990PhR...195..127B,DUBKOV_2008}.  

Levy flights also provide  an appealing playground with  genuine scale-independent developed non-Gaussianities (that cannot be mimicked by a mere local nonlinear transformation of the fields), whose amplitudes  resemble what is expected in gravitational density fields. These remarks would be of limited interest however if we had no way of exploring the statistical properties of such fields. Building on the early results of \cite{1980lssu.book.....P} and of 
\cite{1999A&A...349..697B},
\cite{2022A&A...663A.124B} derived a large corpus of properties of such fields. They were concerned to a large extend on the joint density PDFs at large separation. 

In this paper, we focus on the local behaviour of the field, trying to first grasp the expected behaviour of the local density and its derivatives, so as  to identify critical points in such density field.
The starting point of those investigations  is based on the derivation of the cumulant generating function (CGF) in multiple cells. 
 This allows us to predict their one and two point statistics   to arbitrary order in the variance of the field.
For clarity, the main text focuses on results, while all the derivations are given in the Appendices.

\begin{figure}
\includegraphics[width=1.\columnwidth]{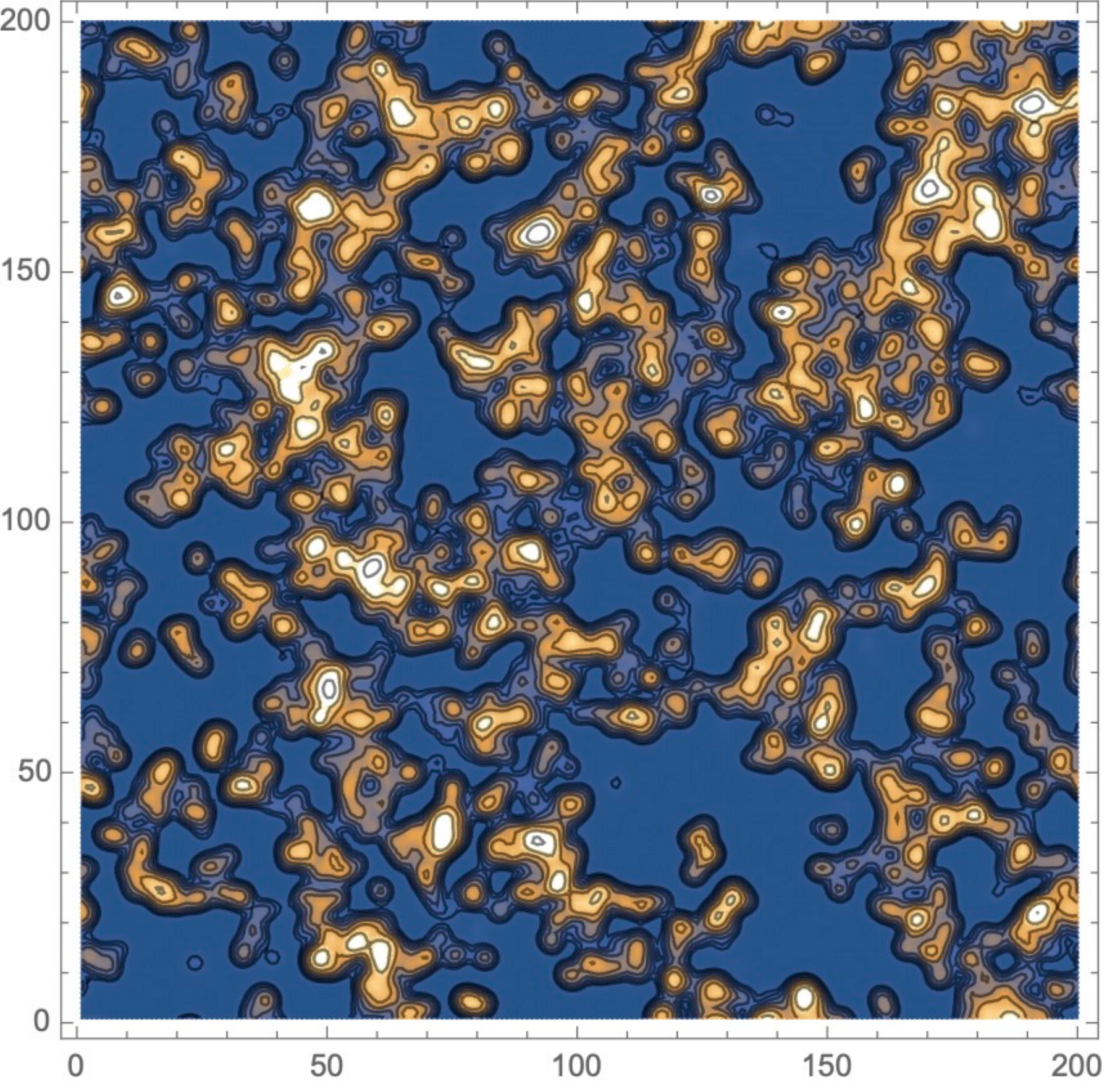}
\caption{Example of a 2D density field derived from a 2D levy flight. Parameters of the flight are $\alpha=1.$, $\ell_0=0.012$ pixel size with 10 points per pixel. The sample has periodic boundary conditions with $200^2$ pixels. The field has been convolved with a Gaussian window function of width 2 (pixel size) in each direction. The resulting variance as measured in the sample is 1.56. On the plot, the contour lines are log-spaced, from density of about 0.2 to 7. The deep blue regions correspond to empty regions.}
\label{fig:2Drealization}
\end{figure}

Section~\ref{sec:levy} recalls the main relevant properties of \flights\!. 
Section~\ref{sec:numbercount} first presents the number count of extrema of  flights in 1,2 and 3D dimensions,
while  Section~\ref{sec:ExtremaCorr} compute their clustering.
Comparison  to Monte Carlo simulation are provided throughout for validation for the 1D flights.
The companion paper will provide them  for 2 and 3D flights.
 Section~\ref{sec:conclusion} wraps up.

\section{The Rayleigh Levy flight model}
\label{sec:levy}

The  random  \flights model, as introduced first in a cosmological context by J. Peebles \citep{1980lssu.book.....P}, is defined as a Markov random walk where the PDF of the step length $\ell$ follows a cumulative cumulative distribution function given by
\begin{subequations}
    \begin{align}
P(>\ell_{0})&=1,\\
P(>\ell)&=\left(\frac{\ell_{0}}{\ell}\right)^{\alpha}\,,\ \ \hbox{for}\ \ \ell>\ell_{0}, 
\label{eq:defPell}
\end{align}
\end{subequations}
where $\ell_{0}$ is a small-scale regularization parameter. 
The random walks are dominated by rare, large events rather than the accumulation of many steps.

Calling $f_{1}(\vr)$  density of the subsequent point (the first descendant) at position $\vr$,
\begin{equation}
f_{1}(\vr)=\frac{\alpha \Gamma(D/2+1)}{2 \pi^{D/2}}\frac{\ell_{0}^{\alpha}}{\vert\vr-\vr_{0}\vert^{D+\alpha}},\ \ \hbox{in D-dimension space},
\end{equation}
and $\psi(k)$ it's Fourier transform
\begin{equation}
\psi(k)=\int\dd^{D}\vr\ f_{1}(\vr)\,e^{-\ii\vk \cdot\vr}, \label{eq:defpsi}
\end{equation}
 the two-point correlation functions for  \flights between positions $\vr_{1}$ and $\vr_{2}$ is given by
\begin{equation}
\xi(\vr_{1},\vr_{2})
=\frac{1}{n} \int\frac{\dd^{D}\vk}{(2\pi)^{D}}\ e^{\ii\vk \cdot (\vr_{2}-\vr_{1})}\,\frac{2}{1-\psi(k)}\,,\label{exactxiRL}
\end{equation}
where $n$ is the number density of points in the sample.
When $k l_0\ll 1$ then $1-\psi(k) \sim (\ell_0 k)^\alpha$  the large distance correlation function behaves like
\begin{equation}
\xi(\vr_{1},\vr_{2})
\sim r^{\alpha-D}\ell_0^{-\alpha}/n.
\end{equation}
The expected behaviour of the large-distance correlation function is  shown on Fig. \ref{fig:RLflights_xi1D} below, which clearly shows that this form is valid as soon as $r\ge \ell_0$.

\begin{figure}
\includegraphics[width=0.99\columnwidth]{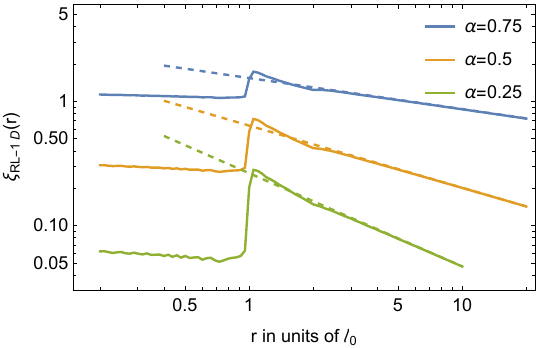}
\caption{The two-point correlation function for the 1D \flights (before smoothing). The predictions are for $\ell_0=1$. The solid lines are the exact shapes derived from equation~\eqref{exactxiRL} exhibiting a clear exclusion zone for $r<\ell_0$. 
The dashed is the large scale asymptotic form. it can be observed that the exact solution converges very rapidly to the asymptotic form.}
\label{fig:RLflights_xi1D}
\end{figure}

Note that the numerical system is driven by two parameters $\ell_0$ and $n$, with $\xi_0\sim 1/(n \ell_0^\alpha)$. In practice one wants to make sure that the filtering scale is significantly larger than both $\ell_0$ and $\ell_n\equiv 1/n^{1/D}$ and that the survey size, $\ellS$, is much larger than the filtering scale. When measuring the correlation properties of critical points, we also want to make sure that $\ellS$ is much larger than the
scales at which the correlation functions are measured. It is always possible to meet these requirements, and reach any amplitude of $\xi$, provided one has enough resources to build \flights with large number of points.
{
In practice, in the 1D numerical simulations we exploited in the following we have 
\begin{center}
\begin{tabular}{ccc}
  \hline
 \small  $\alpha$ & = &  \small 0.5 \\
    \hline
  \small   $n$ & = &  \small 3 / pixel\\
     \hline
   \small  $\ell_0$ & = & \small 0.02 pixel\\
     \hline
   \small  $r_G$ & = & \small 2 pixel\\
     \hline    
    \small  $\ellS$ & = & \small $10^5$ pixel size\\
     \hline
   \small  $N_{\rm realizations}$ & = & \small 200\\
      \hline
\end{tabular}
\end{center}
providing very accurate determination of all one-point quantities (density and critical points, etc.) and good estimates of their correlation as shown subsequently.

A key property of \flights involves the structure of the higher order correlation functions. In such a model, not only are they known but they also display the type of behaviour one expects in the context of cosmological models. We know from theory in the perturbative regime \citep{Review2002}, and to a large extend from observations \citep{2022A&C....4100662Y} that the $p$-order matter correlation functions scale like the power $p-1$ of the two point function. 
This assumption and its consequences  are addressed in more details in the Appendices~\ref{sec:model}-\ref{sec:MTM}. 

Conversely, from a theoretical perspective, the \flights model falls within the  class of models whose $p$-point correlation 
functions follow the so-called hierarchical Ansatz \citep{1984ApJ...279..499F}, so that the 
they can be expressed as a sum of products of two-points correlations
\begin{equation}
\xi_{p}(\vr_{1},\dots,\vr_{p})=\sum_{t\in\trees}Q_p(t)\,\prod_{\lines\in t}\xi(\vr_{i},\vr_{j})\,,
\label{eq:defhansatzmaintext}
\end{equation}
This process actually embodies a more specific class of models,  which correspond to cases where the $Q_p(t)$ values depend only on the vertex composition of the tree, that is
\begin{equation}
Q_{p}(t)=\prod_{{\rm vertices}\in t}\nu_{q}\,,
\label{eq:deftreemaintext}
\end{equation}
where the vertices values $\nu_q$ are numbers associated with each vertex and depend only on their connectivity, $q$ being here the number of lines a given vertex is connected to,  \citep[see Appendix~\ref{sec:HTM} and][]{1999A&A...349..697B}.
In the case of the flight models we consider here, we have only two non-zero vertices, $\nu_1$ and $\nu_2$ so that the trees we have to consider are in fact simply connected lines. 

In general, the corresponding  cumulant generating function of the underlying density field in cells of arbitrary profile given by $\mW_{i}(\vx)$ can be written as 
\begin{equation}
\varphi(\lambda_{1},\dots,\lambda_{n})
= {\hskip -0.2cm}
\sum_{p_{1},\dots,p_{n}}\langle\rho_{1}^{p_{1}}\dots\rho_{n}^{p_{n}}\rangle_{c}\,\frac{\lambda_{1}^{p_{1}}}{p_{1}!}\,.
\label{def:CGF}
\end{equation}
Note that in the Gaussian limit this expression is restricted to terms with $p_1+\dots+p_n=2$. Note also that in general the full complexity of this expression is hard to exploit and can only be done in a perturbative sense, exploring  for instance the consequences of terms at cubic order. This is at the heart of the Edgeworth expansion \citep{1977ats..book.....K,2017arXiv170903452S}.

A remarkable property of the tree models in general, and of the Levy flight model in particular,  is that the discrete summations that appear in this expression (over $p_1, \dots, p_n$), can be done explicitly. This is clearly an appealing feature of these models,
as it allows for an exploration of their properties in a regime that can be arbitrarily far  from the Gaussian case. This is the motivation for this study.

The details of the derivation of the resulting form for the CGF, $\varphi(\lambda_{1},\dots,\lambda_{n})$, is given in  Appendix~\ref{sec:HTM}, which shows that 
\begin{align}
\varphi(\lambda_{1},\dots,\lambda_{n})
&=\sum_{j}\lambda_{j}
\int{\dd\vx}\,\mW_{j}(\vx)\left(1+{\tau(\vx)}/{2}\right)\,,
\label{treesumphi}
\end{align}
where    $\tau(\vx)$ is the $\lambda_i$-dependent implicit solution of  the consistency equation
\begin{equation}
\tau(\vx)=\sum_{j}\lambda_{j}\int{\dd\vx'}\,\mW_{j}(\vx')\,\xi(\vx,\vx')\,\left(1+{\tau(\vx')}/{2}\right)\label{treesumtau}.
\end{equation}
Formally, $\tau(\vx)$ is actually the rooted-Cumulant Generating Function (the generating function of cumulants that originate from location $\vx$) and is denoted r-CGF hereafter. It will play an important role throughout this paper. Expression~\eqref{treesumtau} is the direct transcription of the tree model described by equation~\eqref{eq:deftreemaintext}. It is the result of combinatoric computations, and does not rely on any approximation\footnote{
Yet, it is probably inaccurate to assume that  flights with a finite number of points are equivalent to Poisson realisations drawn from a continuous field whose   correlations are computed in the continuous limit.}.
The practical implementation of equation~\eqref{treesumphi} relies on some approximations, such as the  mean field assumption, in which the r-CGF is assumed to remain constant within each cell (as detailed in Appendix~\ref{sec:HTM}). This proves to be highly accurate for compact, non-compensated, spherically symmetric density profiles. Assuming  it holds,
let us denote $\tau_i$ the value of the r-CGF for each cell $i$. We can then average equation~\eqref{treesumtau} over the cell $i$ with  profile $\mW_i(\vx)$, which leads to the following system
\begin{equation}
\tau_i=\sum_{j}\lambda_{j}\,\left(1+{\tau_j}/{2}\right)\int{\dd\vx}\int{\dd\vx'}\,\mW_{i}(\vx)\,\mW_{j}(\vx')\,\xi(\vx,\vx')\,,\label{treesumtaudiscrete}
\end{equation}
which becomes a set of $n$ equations coupling the different values of $\tau_i$. As can be seen here, in case of the present minimal tree model, this system is linear in $\tau_i$. It can therefore be explicitly 
inverted\footnote{For more general tree models, the system is fully nonlinear and the inversion is done numerically.}. 
For the one-point cumulant generating function, equation~\eqref{treesumtaudiscrete}
corresponds to the implicit equation
\begin{equation}
\tau_1=\lambda_1\xi_0\left(1-\tau_1/2\right)\,,\label{treesumtaudiscrete1D}
\end{equation}
where $\xi_0$ is the average two-point correlation function within one cell
\begin{equation}
\xi_0=\int{\dd\vx}\,{\dd\vx'}\,\mW_{1}(\vx)\,\mW_{1}(\vx')\,\xi(\vx,\vx')\,\label{def:xi0}\,.
\end{equation}
The resulting CGF can easily be computed (See Appendix~\ref{sec:MTM}):
\begin{equation}
\varphi_{1}(\lambda_{\rho};\xi_{0})=\frac{2 \lambda _{\rho }}{2-\xi _0 \lambda _{\rho }}\,.
\label{varphi1}
\end{equation}
Despite its apparent simplicity, this expression fully captures the whole cumulant hierarchy that one expects for the density PDF. Furthermore, it turns out that the inverse Laplace transform of such an expression can be explicitly derived, see \cite{2022A&A...663A.124B} and appendix~\ref{sec:MTM}, and leads to the closed form expression of the density PDF
\begin{align}
P(\rho)&=\int_{-\ii\infty}^{+\ii\infty}\frac{\dd \lambda_{\rho}}{2\pi\ii}\exp\left(-\lambda_{\rho}\,\rho+\varphi_{1}(\lambda_{\rho})\right)\,,\label{eq:PDFgeneric}
\\
&=e^{-2/\xi _0}\Dirac(\rho)+\frac{4} {\xi _0^2}e^{-\frac{2 (\rho +1)}{\xi _0}} \, \tHyp\left(2;\frac{4 \rho }{\xi _0^2}\right)\,. \label{eq:P1Dexplicit}
\end{align}
The first (singular) term in equation~\eqref{eq:P1Dexplicit} reflects the contribution to empty regions.  
The last terms involves the the regularized confluent hypergeometric function,  $\tHyp(a;z)$, given by $x^{\frac{1-a}{2}} I_{a-1}\left(2  \sqrt{x}\right)$.
Note that this model indeed predicts region that are empty even for an arbitrarily large number of points\footnote{For a growing number of points  at fixed $l_0$, the variance is decreasing and the VPDF will eventually vanish. If however the number density of points increases while $n l_0^\alpha$ is fixed, ensuring $\xi_0$ is fixed, the VPDF  remains finite.}.  
More specifically, the void probability function (VPDF) is non zero, even in the continuous limit and is given by \begin{equation}
    P_0=e^{-2/\xi_0}\,.
\end{equation}
This feature can be appreciated on Fig.~\ref{fig:2Drealization}, where one can see empty regions that covers a large fraction of the sample. {The VPDF can only be non-zero for a compact support filter, which not formally holds for a Gaussian filter. Yet the mean field result seems to accurately predicts the VPDF for a top-hat filter. For a Gaussian filter the behaviour of the density PDF in the low density regime is correct when corrections to the mean field solutions are included.}

Note that for the mean-field solution, the dependence on the space's dimension, the value of $\alpha$ and the filter shape is entirely contained in the value of $\xi_0$. This is not the case for  derivations of the density PDF  beyond the mean-field solution, which are expected to depend on the details of the model and filtering schemes. Numerical investigation beyond the mean field  
show  nonetheless that the high density tail of  equation~\eqref{eq:P1Dexplicit} is very robust. This is illustrated on Fig.~\ref{fig:PDFMFExpand},
which displays  the mean-field solution and its corrections in terms of Hermite polynomials (up to sixth order in the expression of the r-CGF, see Appendix~\ref{sec:MTM}),
and compares it to numerical results.

\begin{figure}
\includegraphics[width=1.\columnwidth]{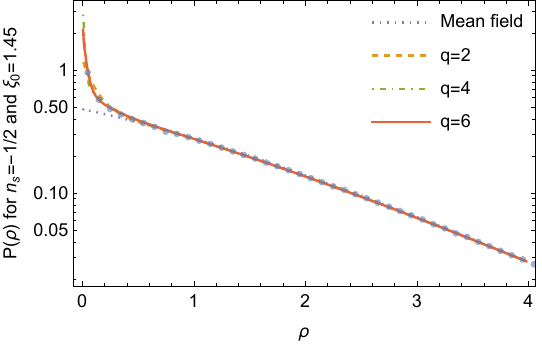}
\caption{Comparisons of measured density PDF from a set of 1D  \flights whose characteristics are given in the text, grey points, compared to different levels of theoretical predictions. The blue dotted line is the mean field approximation. The other lines correspond to different level of approximation, up to sixth order in an expansion in Hermite Polynomials. 
}
\label{fig:PDFMFExpand}
\end{figure}

\section{Critical point number counts}
\label{sec:numbercount}
\begin{figure*}
\includegraphics[width=0.99\columnwidth]{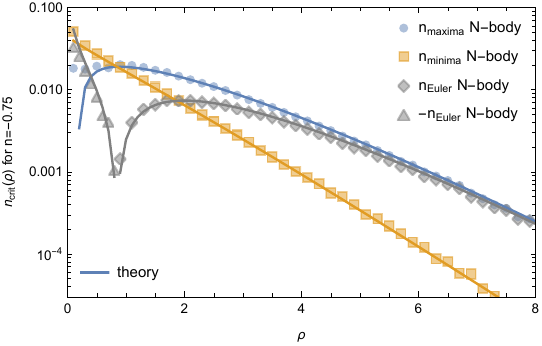}
\includegraphics[width=0.99\columnwidth]{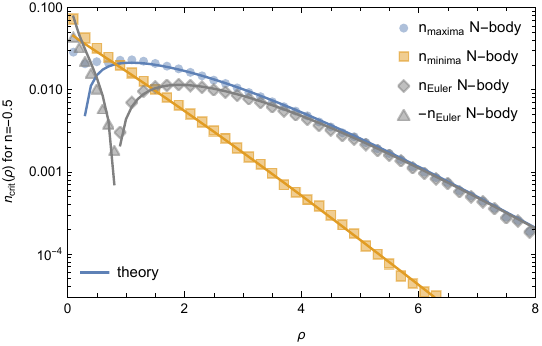}
\caption{Mean field prediction for extrema counts and comparison with 1D numerical Levy flight for two different indices as labelled. See text for details on the measurements. Note the Euler number density is negative on the low density branch as the number density of minima exceeds the number density of maxima.
The agreement between theory and measurements is remarkable. 
}
\label{fig:Extrema1D}
\end{figure*}
From a given realisation of the flight, a convolution by e.g. a Gaussian filter allows us to define a 
smooth field, $\rho$, whose critical point can be studied statistically. 
The number density of critical points at density $\rho$ then reads 
\begin{equation}
    n_{\rm crit.}(\rho)\!=\!
\!\!\int\!\!\prod_{ij}{\dd\rho_{\!,ij}}
  {\rm Sgn}\left[\rho_{\!,ij}\right]\! \det\left[\rho_{\!,ij}\right]\!
    P\!\left(\rho,\rho_{\!,i}=0,\rho_{\!,ij}\right),  \label{eq:defeuler}
\end{equation}
where ${\rm Sgn}\left[\rho_{,ij}\right]$
is either $0$, $1$ or $-1$, depending on the sign of the eigenvalues of the matrix $\rho_{,ij}$, to reflect the nature of the critical points considered. 
Hence the computation of extrema densities requires the joint cumulant generating function of variables dual to the local density, its first and second order derivatives. 
Appendix~\ref{sec:ExtremaDerivation} shows  that it is indeed possible to   derive such joint CGF in the mean field limit. 
In short, the calculation is based on writing the mean field solution for a finite number of cells assumed to be infinitely close to one another; the derivatives are then obtained via finite differences.

 Once the cumulant generating function for the field and its derivatives  is known,  the relevant conditional expectations are computed to predict the extrema and critical number counts and their clustering properties.  %
We present below the results for the Euler and critical points in  dimensions one to three. 
As equation~\eqref{eq:defeuler} involves up to the second derivative of the field, in principle one should compute the joint PDF of the field and its derivative up to that order. 
In practice, For 1D Euler number counts, Appendix~\ref{sec:ExtremaDerivation} shows explicitly that thanks to the stationarity of the field, only the first derivative is necessary, and we can write
\begin{equation}
    n_{\rm 1D}^{\rm Euler}(\rho)=\frac{\partial}{\partial\rho}\int_{-\infty}^{0}
    \rho_{\!,x}\dd\rho_{\!,x}P(\rho,\rho_{\!,x})\,.
\end{equation}

Overall, and after a significant amount of algebra, we derive in Appendix~\ref{sec:ExtremaDerivation}    the closed form Euler counts in ND: 
\begin{eqnarray}
n_{\rm ND}^{\rm Euler}(\rho)&=& \frac{1 }{
   \xi _0^{2} }
   \left(
   {\frac{2\xi_1}{\pi \xi _0^{2}\rho }}  \right)^{N/2}
e^{-\frac{2 (\rho +1)}{\xi _0}}  \nonumber\\
&& \hspace{-2.2cm}\times  \left\{
P_{N-1}( \xi _0,\rho) \, \tHyp\left(1;\frac{4 \rho }{\xi
   _0^2}\right)+
  Q_{N}( \xi _0,\rho) \, \tHyp\left(2;\frac{4 \rho }{\xi _0^2}\right)
  \right\}\,, \label{eq:EulerNDflight}
\end{eqnarray}
where $P_m$ and $Q_m$ are polynomials of order $m$ in 
$\rho$ and $\xi_1$ is defined in equation~\eqref{xi0def}. Specifically we find in Appendix~\ref{sec:ExtremaDerivation} that 
\begin{subequations}
\begin{eqnarray}
    P_0&=&2 \xi_0,\quad
P_1 =\!-\xi_0 (\xi_0+8 \rho)/2, \\
P_2 &=&\!\left(16 \xi _0 \rho  (3 \rho +1)\!+\!3 \xi _0^3\right)/8, \\
 Q_1&=&\!\!\!-(\xi_0+4 \rho),\,\,
Q_2=(2 \xi_0 \rho +\xi_0^2+ 8 \rho (\rho+1))/2,\\
Q_3&=&\!\!\!-\left(6 \xi _0^2 \rho +16 \xi _0 \rho +3 \xi _0^3+32 \rho ^2 (\rho +3)\right)/8 . 
\end{eqnarray} \label{eq:defPQ}
\end{subequations}
Here, $
   \xib_\mathrm{D}\!\!\propto \ell_0^{-\alpha } r_s^{\alpha -D}
$
where the pre-factor defining $\xi_0$ is a function of $\alpha$ only, given by equations~\eqref{axiRL1D}-\eqref{axiRL3D} for 
dimension 1 to 3.

The simplicity of equation~\eqref{eq:EulerNDflight} is quite remarkable. Note in particular that 
for $D=2$ to 3, $n_{\rm ND}^{\rm Euler}(\rho)$ also does not involve $\xi_2$ (see the discussion below and Appendix~\ref{sec:ExtremaDerivation}).

One could speculate that  relations such as equations~\eqref{eq:defPQ} hold in higher dimensions, involving polynomials,
$P_i$ and $Q_i$ of increasing order, so that equation~\eqref{eq:EulerNDflight}
mirrors the hierarchy for Gaussian random field Euler counts  \citep[involving Hermite polynomials,][]{Adler2009}
which follow from their cumulant generating functions given in Appendix~\ref{sec:gauss}.

It is of interest to be able to distinguish between the number counts of maxima and minima separately.
It turns out   that the extrema counts can also be computed via specific (simple or double) numerical integration path in the complex plane as 
\begin{align}
n_{\rm 1D}^{\rm \pm}(\rho)&=
\int_{-\ii\infty\pm\eps}^{+\ii\infty\pm\eps}\frac{\dd\lambda _{xx}}{2\pi\ii}
\frac{1}{\lambda_{xx}^{2}}
\mixchi_{\rm 1D}(\rho;\lambda_{xx})\,,
\label{P1Dextrema}
\end{align}
and 
\begin{equation}
n_{\rm 2D}^{\rm \pm}(\rho)= 
\int_{-\ii\infty\pm\eps}^{+\ii\infty\pm\eps}\frac{\dd\lambda_{\kappa}}{2\pi\ii}
\int_{0}^{\infty}\frac{\dd\lambda_{\gamma}}{2\pi}\nonumber\\
\frac{12\pi\lambda_\gamma}{(\lambda_\gamma^2+\lambda_\kappa^2)^{5/2}}\ 
\mixchi_{\rm 2D}(\rho;\lambda_{\kappa}, \lambda_{\gamma})\,,
\label{P2Dextrema}
\end{equation}
where $\mixchi_{\rm 1D}(\rho;\lambda_{xx})$  and $\mixchi_{\rm 2D}(\rho;\lambda_{\kappa}, \lambda_{\gamma})$ are   given by equations~\eqref{mixchi1D} and ~\eqref{mixchi2D} resp., while $\eps$ is a negative real constant for maxima and a positive real constant for minima. The 2D saddle counts follows from $\sum_\pm n_{\rm 2D}^{\rm \pm}(\rho)-n_{\rm 2D}^{\rm Euler}(\rho)$. 

To gain a bit of insight into these quantities, let us consider 
the expression of $n_{\rm 1D}^{\rm \pm}(\rho)$ as a function of $\gamma= \xi_1/\sqrt{\xi_2 \xi_0}$ in the limit\footnote{Given that $0<\gamma<1$ is  the ratio of the distance between zero crossing and extrema,  it is always possible to add arbitrary large amount of small scale fluctuations, which corresponds to this limit.} $\gamma\to 0$. The integrand in this case is greatly simplified and it leads to the same leading behaviour for the number density of maxima and minima given by
\begin{equation}
    n_{\rm ext}(\rho)\dd\rho=\frac{1}{2\pi}\left(\frac{\xi_2}{\xi_1}\right)^{1/2}P(\rho)\dd\rho\,,
\end{equation}
which, as expected, scales like the inverse of typical distance between extrema, $\left({\xi_1}/{\xi_2}\right)^{1/2}$.
The next to leading order is independent of $\gamma$,
and corresponds to the Euler number density so that
\begin{equation}
    n^\pm_{\rm 1D}(\rho)\dd\rho=\frac{1}{2\pi}\left(\frac{\xi_2}{\xi_1}\right)^{1/2}P(\rho)\dd\rho \pm\frac{1}{2}n^{\rm Euler}_{\rm 1D}(\rho)\dd\rho+\dots
\end{equation}
where dots represents subsequent terms in a $\gamma$ expansion.
{Here  $n_{\rm Euler}$ is less sensitive to small scale high frequency 
features -- or noise -- compared to extrema number density. This is also the case for Gaussian field and for the \flights in any dimensions. Although we cannot demonstrate it, we speculate that this is true for all hierarchical models in which the  high order correlation function behave like products of the two-point correlations.}
\begin{figure}
\includegraphics[width=0.99\columnwidth]{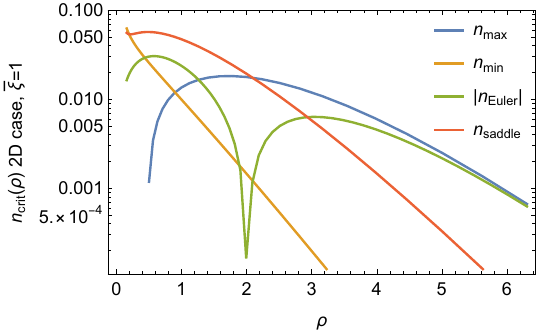}
\caption{Mean field prediction for extrema counts for 2D fields. The plots are for $\xi_0=1$ and $\gamma=0.45$.}
\label{fig:Extrema2D}
\end{figure}
Figs~\ref{fig:Extrema1D} and ~\ref{fig:Extrema2D}  present the corresponding number counts in one and two dimensions resp. In Fig.~\ref{fig:Extrema1D} in particular we present a detailed comparison of the extrema, minima and Euler number densities as a function of the local density. We take  advantage here of the fact that we could rely on large number of realisations making the measurements of the critical points rather precise (numerical error bars due to the finite number of realisations are here negligible). For the 1D case, it is also rather simple to determine the position and type of critical points: they are obtained at locations where the local gradients (obtained after convolution on grid points shifted by 1/2 pixel) change sign from one pixel to the next. The type of critical point is simply given by the sign of the difference.
Fig.~\ref{fig:Extrema1D} shows that the theoretical predictions capture in exquisite details the measured densities of critical points. The only significant departure between the theoretical predictions and the numerical results is for the low density regime ($\rho<1$), for the maxima number densities and to a less extent the minima number densities. These discrepancies are thought to be related to the breaking of the mean field approximation in the low density regime, as shown on Fig.~\ref{fig:PDFMFExpand} for the density PDF. All these conclusions are found to be valid irrespective of the details of the simulations (this is illustrated with a change of the value of $\alpha$ in the \flights\!\!).
\section{Clustering of critical points}
\label{sec:ExtremaCorr}
One of the strength of analytical investigations of  tree-hierarchical models is that the same re-summation techniques can be used to also infer the  large distance correlation functions of the objects for which we can compute the number density. These calculations were pioneered in \cite{1999A&A...349..697B} for tree hierarchical models, applied also to the perturbation theory calculations in \cite{1996A&A...312...11B},  and more recently in \cite{2016MNRAS.460.1598C,2017MNRAS.466.2067U,2018JCAP...01..053M,2021MNRAS.500.3631R}. A general thorough presentation of large-scale biasing is also to be found in the review paper by \cite{2018PhR...733....1D}.

The properties of large scale biasing were more specifically investigated in \cite{2022A&A...663A.124B} for the Rayleigh-Levy model,
which focused on the consequences of this functional form for the covariance properties of density PDF measurements.
These models share the same properties: the join density PDF at large separation are expected to obey the following functional form
\begin{equation}
    P(\rho_1,\rho_2)=P(\rho_1)P(\rho_2)
    \left[1+\xi(r_1,r_2)b(\rho_1)b(\rho_2)\right]\,,
    \label{densitybias}
\end{equation}
where the densities $\rho_1$ and $\rho_2$ are taken at position $r_1$ and $r_2$ respectively. At this order, the dependence in the densities $\rho_1$ and $\rho_2$ factorises, making it possible to define a density bias function.  In this paper, we present how such relation can be extended to number counts of \textit{critical points} and derive the corresponding bias functions.

In equation~\eqref{densitybias} the bias function can be seen as the response function of the density PDFs to a change of the global density.
This means that although the bias function cannot be derived from the density PDFs alone, we should be able to derive it if an operational method is available to compute the density PDFs for arbitrary large-scale density (following the derivation of halo-bias function as pioneered by \cite{1996MNRAS.282..347M} in a separate universe approach).
In  hierarchical tree models, it can be computed making use of the r-CGF. The shape of the join density CGF at for densities at positions $r_1$ and $r_2$ written up to first order in $\xi(r_1,r_2)/\xi_0$ indeed obeys
\begin{equation}
\varphi(\lambda_1,\lambda_2)
=
\varphi_1(\lambda_1) \varphi_1(\lambda_2)+
\tau_1(\lambda_1)\xi(r_1,r_2) \tau_1(\lambda_2)\,,
\label{eq:phixiexpansion}
\end{equation}
where $\tau_1$ is the r-CGF introduced in equation~\eqref{treesumtau}. 
The computation of the inverse Laplace transform of such a join CGF
can be formally done at leading order in $\xi(r_1,r_2)$ and has the functional form given by equation~\eqref{densitybias}.
More specifically, equation~\eqref{eq:PDFgeneric} is to be extended to
\begin{equation}
    b(\rho)P(\rho)=\int_{-\ii\infty}^{+\ii\infty}\frac{\dd \lambda_{\rho}}{2\pi\ii}
    \tau_{1}(\lambda_{\rho})
    \exp\left(-\lambda_{\rho}\,\rho+\varphi_{1}(\lambda_{\rho})\right)\,\\
\label{eq:bPDFgeneric}
\end{equation}
for the bias function. For the \flights model, the large scale density bias function can be derived explicitly \citep{2022A&A...663A.124B} and it reads
\begin{equation}
    b(\rho)=\frac{\tHyp\left(1;\frac{4 \rho }{\xi _0^2}\right)}{\tHyp\left(2;\frac{4 \rho }{\xi _0^2}\right)}-\frac{2}{\xi_0}.\\
\label{eq:bexplicit}
\end{equation}
For large density,  its limiting behaviour  is
\begin{equation}
    b_{\rm large}(\rho)=\frac{1}{4}+\frac{2}{\xi_0}\left(\sqrt{\rho}-1\right),
\label{eq:blarge}
\end{equation}
which corresponds  to a dependence in $\sqrt{\rho}/\xi_0$ for large densities.
Note that this is not a priori a generic result for hierarchical models and is at variance with the behaviour found in the context of Perturbation Theory calculations in \cite{2016MNRAS.460.1598C} where the bias at large density is expected to be proportional to the density\footnote{To be more precise, the \flights model lead to identical CGF and r-CGF,  whereas these two functions are expected to exhibit different singularities on the real axis for generic tree model, as shown by \cite{1999A&A...349..697B}.}. 

The objective of this section is to present the correlation properties of the critical points of \flights\!\!~.
Correlation properties of critical points will depend in general on 
  i)  the type of critical points one considers;
   ii) the distance between the points and how this distance compares to the scale at which these points are defined. 
Despite the fact that the model we consider is unambiguously defined, it is near impossible to derive the correlation properties of critical points in their full complexity! Results can be derived at large enough separation and using the mean field approach, implying that our findings  will be solid for large enough densities only.

The starting point is a generalisation of equation~\eqref{eq:phixiexpansion} for the joint CGF of the density and its derivatives,
\begin{align}
    \varphi\left(\lambda_{\rho},\!\{\lambda_{i}\},\{\lambda_{ij}\};\mu_{\rho},\!\{\mu_{i}\},\{\mu_{ij}\}\right)
  & \!=\! \varphi\left(\lambda_{\rho},\!\{\lambda_{i}\},\{\lambda_{ij}\}\right)\!+\!
    \varphi\left(\mu_{\rho},\!\{\mu_{i}\},\{\mu_{ij}\}\right)\nonumber\\
&\hspace{-3cm}+   
\tau\left(\lambda_{\rho},\{\lambda_{i}\},\{\lambda_{ij}\}\right)\,\xi_0(d) \,
    \tau\left(\mu_{\rho},\{\mu_{i}\},\{\mu_{ij}\}\right)\,, \label{eq:defvarphimultiple}
\end{align}
where $\lambda_{\rho},\{\lambda_{i}\},\{\lambda_{ij}\}$ are the conjugate variables of the density and its derivatives at a given location, 
$\mu_{\rho},\{\mu_{i}\},\{\mu_{ij}\}$ those at a second location placed at distance $d$ from one another. $\xi_0(d)$ is the filtered matter density correlation function at distance $d$. This expansion is valid when $d$ is much larger than the filtering scale and derived from the fact that we then expect
$\xi_0(d)\ll\xi_0$ and $\xi_2(d)/\xi_2\ll\xi_0(d)/\xi_0$ (see Appendix~\ref{sec:extremaCorrelation}). 
For the Rayleigh-Levy model we further have $\tau=\phi$. 
As for the density field this form implies a similar form for the two-point number density of the critical points
\begin{align}
 n_{\#1,\#2}(\rho_1;\rho_2)&\!=\!  n_{\#1}(\rho_1)
    n_{\#2}(\rho_1)\!
    \left\{
    1\!+\!b_{\#1}(\rho_1)\xi_0(d)b_{\#2}(\rho_2)
    \right\}\!, \label{eq:n1n2fact}
\end{align}
where $\#i$ represents the types of critical points one consider and $b_{\#i}(\rho)$ is the associated bias function whose expression is given by \eqref{eq:bcritformal}.

The computation of bias function of the critical points then makes use of the same techniques as for the number density. It does not lead to an explicit form for the extrema or saddle point position both in 1D and 2D. It is however possible to derive the explicit bias functions of the Euler points for the 1D and 2D case:
 \begin{align}
b_{\rm Euler}^{\rm ND}(\rho) &n_{\rm Euler}^{\rm ND}(\rho)
=
\frac{1}{\xi_0^3}\left(\frac{2\xi_1}{\pi  \xi _0^2 \rho }\right)^{N/2} e^{-\frac{2 (\rho +1)}{\xi _0}} 
\nonumber\\
& \hskip -.9cm\times
\left\{{\cal P}_N(\xi_0,\rho) \,
   \tHyp\left(1;\frac{4 \rho }{\xi _0^2}\right)
  +
   {\cal Q}_N(\xi_0,\rho) 
   \, \tHyp\left(2;\frac{4 \rho }{\xi _0^2}\right)\right\}. \label{eq:bresults}
\end{align}
where the polynomials obey 
\begin{subequations}
\begin{align}
{\cal P}_1&=\xi _0\,(\xi_0 \!-\! 4 (1 \!+\! \rho)), \quad
{\cal Q}_1=2 (\xi_0 + 8 \rho), 
\\
{\cal P}_2&\!=\!\xi_0 (\xi_0 \!-\! 3 \xi_0 \rho \!+\! 4 \rho (3 \!+\! \rho)),
\,\,
{\cal Q}_2\!=\!-  \left(\xi_0^2 + 8 \rho (1 + 3 \rho)\right).
\end{align}
\end{subequations}
Again one could conjecture that higher order polynomials exist for bias functions in higher dimensions.
Note that bias functions in general are dimensionless quantities. They  can therefore
be expressed in terms of dimensionless quantities
such as $\xi_0$ and $\gamma$. Furthermore the bias parameter of Euler points is found to be also independent of $\xi_2$ (as was the case for its number density) and therefore of $\gamma$.
%
\begin{figure}
\includegraphics[width=0.99\columnwidth]{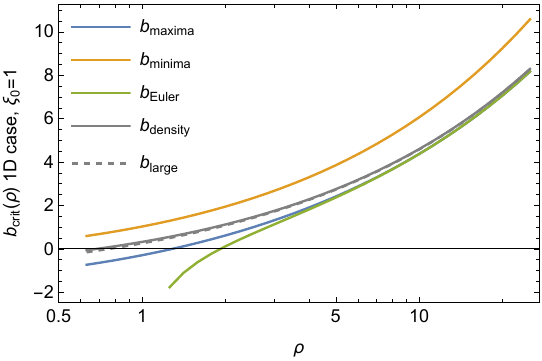}
\caption{Theoretical prediction for the bias function of 1D extrema, of Euler points and that of the local density, \eqref{eq:bexplicit}, and its corresponding large scale limit (dashed line) as given by \eqref{eq:blarge}. As expected, the maxima and Euler points have the same bias at large density, this is also the bias of the density field itself: high density critical points are likely to be maxima of the fields and their correlation properties is to a large extent determine by density bias.}
\label{fig:bextrema1D}
\end{figure}
\begin{figure}
\includegraphics[width=0.99\columnwidth]{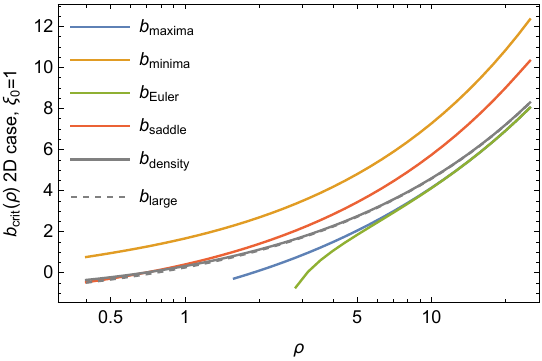}
\caption{Theoretical prediction for the bias function of 2D extrema, saddle points, Euler points and that of the local density (for which the predictions is the same as in 1D). Similar conclusions can be drawn from the 1D case.}
\label{fig:bextrema2D}
\end{figure}

The corresponding bias functions are shown in Fig.~\ref{fig:bextrema1D} and~\ref{fig:bextrema2D} in one and two dimensions resp.  These functions are computed for $\xi_0=1$ and
for $\gamma=0.45$, with corresponds roughly what is expected for $\alpha=0.5$. The results however do not depend crucially on these specific choices.

They exhibit some expected generic behaviours: 
  i)  the large density asymptotic behaviour of maxima and and Euler points are the same. It also matches the asymptotic form of the local density bias: large density regions tend to trace the locations of the maxima. 
   ii) for a given density, minima tend to be more correlated than saddle points or maxima: imposing to have a minima within a large density region indeed requires  even larger density in the surroundings, which in turn translates into  larger biases. 
One important conclusion that can be drawn here is that peak correlation properties derive, to a large extent, for the behaviour of the density bias functions in the sense of equation~\eqref{densitybias}. This justifies to use such a proxy for the computation of peak correlation in the large density regime for more involved models. 
Such an approach however cannot capture proximity or exclusion effects one expects for peak correlations. 

To explore specifically these effects, we rely here on results from numerical experiments of 1D \flights\!\!, for which accurate measurements can be achieved. 
The corresponding curves are presented on  Figs~\ref{fig:xi_clipped_1D_p5} to \ref{fig:xi_cross_MinMax_1D_p5}.
To avoid too much numerical noise and uncertainties, they are expressed in terms of correlation function of cumulative quantities,
 where the local density is above a given threshold.
More precisely the bias factors we will apply are defined as
\begin{equation}
    b_{\#}(>\rho_s)=\frac{\int_{\rho_s}^\infty \dd\rho\ b_{\#}(\rho)n_{\#}(\rho)}
    {\int_{\rho_s}^\infty \dd\rho\ n_{\#}(\rho)}.
\end{equation}
The 1D correlation functions are directly computed in position space by applying a simple shift to the density fields. No binning is applied and they are measured to a separation of 1000 pixels, which is safely smaller than the sample size. 

Fig.~\ref{fig:xi_clipped_1D_p5} shows the correlation functions of the thresholded density for different thresholds. The dashed lines are the theoretical predictions,
i.e. the expectation when one multiplies the measured matter correlation function of the filtered density field (grey solid line) with $b(>\rho_s)^2$. One can see that proximity effects tend to be rather insignificant for low density threshold,
 but rather large for high density threshold: for $\rho_s=4$ proximity effects can be detected up to a distance of about 100 pixels.

We can also predict the actual shape of the two-point function for  \flights models. 
This is presented in Fig.~\ref{fig:RLflights_xi1D}.

Figs~\ref{fig:xi_clipped_1D_p5}-\ref{fig:xi_Euler_1D_p5} displays their corresponding clipped 
two-point functions, while Fig.~\ref{fig:xi_cross_MinMax_1D_p5} shows the cross correlation of 
minima and maxima. 
The consistency of the results with the theoretical predictions show that, on large scales, the functional form of the peak biases, equation~\eqref{eq:n1n2fact},
is indeed satisfied, and the auto- and cross correlation properties of critical points can be described with such a factorized form. 

\begin{figure}
\includegraphics[width=0.99\columnwidth]{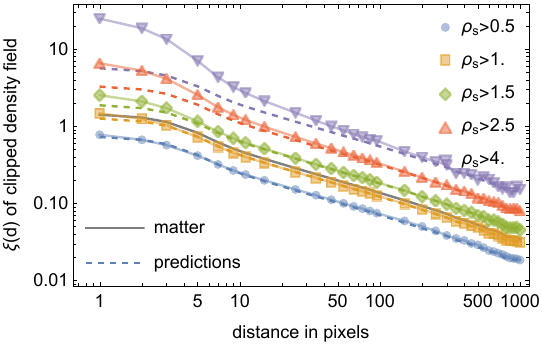}
\caption{The correlation function of the thresholded density regions and comparisons with predictions. The black solid line is the measured matter correlation function. The dashed lines are the prediction correlation amplitudes for thresholded regions
derived from their large-scale limit. They have been computed after applying bias factors $b_\#(>\rho_s)^2$ to $\xi_{0}(d)$ as measured in the simulation. }
\label{fig:xi_clipped_1D_p5}
\end{figure}

\begin{figure}
\includegraphics[width=0.99\columnwidth]{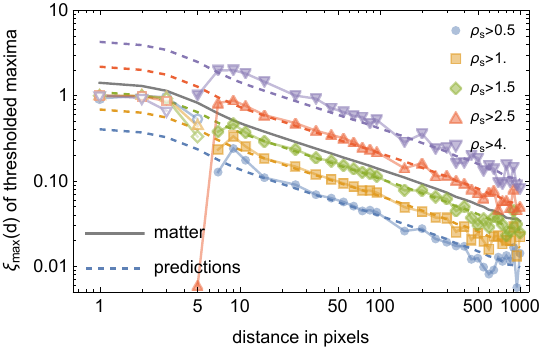}
\caption{Same as previous figure for the thresholded maxima. Open symbols correspond to negative values. We see a sharp transition between the large-scale behaviour -- well predicted by the theory -- and the small scale behaviour. Note that the plateau at small scale corresponds to $\xi_{\rm max}(d)=-1$ meaning that peaks generate an exclusion zone in their vicinity.}
\label{fig:xi_Maxima_1D_p5}
\end{figure}

\begin{figure}
\includegraphics[width=0.99\columnwidth]{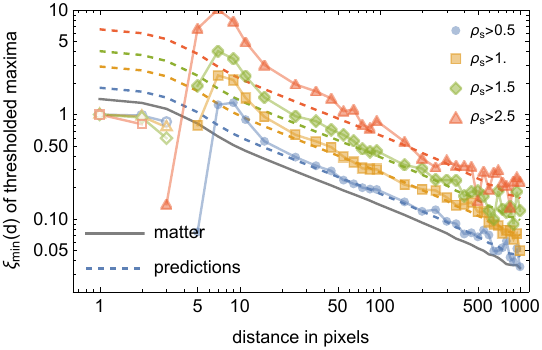}
\caption{Same as previous figure for the thresholded minima. Note that minima are, as expected, more clustered than maxima for a given threshold value. We still observe a transition towards the small scale to $\xi_{\rm maxima}(d)=-1$. Note that this measure is too noisy for a threshold $\rho_s > 4.$ (not shown here).}
\label{fig:xi_Minima_1D_p5}
\end{figure}

\begin{figure}
\includegraphics[width=0.99\columnwidth]{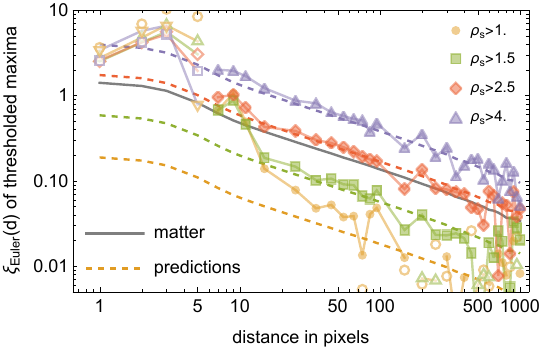}
\caption{Same as previous figure for the thresholded Euler number densities. We still observe a transition towards the small scale. Note that here $\xi_{\rm Euler}(d)<-1$ showing that minima and maxima tend to be correlated at small distance (as confirmed in the next plot). Note that this measure is quite noisy as the Euler number density vanishes for low thresholded values and the threshold $\rho_s > 0.5$ is not shown.}
\label{fig:xi_Euler_1D_p5}
\end{figure}

\begin{figure}
\includegraphics[width=0.99\columnwidth]{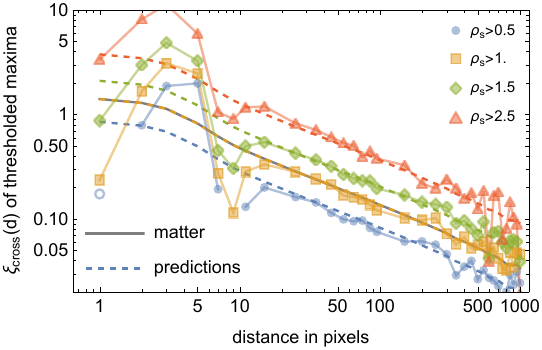}
\caption{Cross correlation between minima and maxima with the same threshold. The prediction is here obtained by multiplying $\xi_0(d)$ as measured in the simulation by $b_{\rm min}(>\rho_s)b_{\rm max}(>\rho_s)$.
We note a large positive correlation for scales that are of the order of the smoothing scale.}
\label{fig:xi_cross_MinMax_1D_p5}
\end{figure}

\section{Discussion and conclusions}
\label{sec:conclusion}

Rayleigh-Levy flights can serve as numerical models portraying matter distribution within highly nonlinear fields in cosmology. In contrast to standard Markovian processes, Rayleigh-Levy flights display long-range correlations across all orders, which are precisely understood. Thus, these models serve as simplified representations, capturing statistical properties of the cosmic density field after gravitational instabilities reach its full development, although we are well aware they do not precisely replicate the outcomes of these instabilities. Despite this, they exhibit key properties, such as hierarchical structure in the higher correlation functions, making their study a valuable reference point.

Let us emphasise that Rayleigh-Levy flights differ from simple nonlinear transformations \citep[like a lognormal fields  introduced by][which are now widely used as toy models]{1991MNRAS.248....1C}, as they exhibit hierarchical properties across scales. This makes them the only known example (at least to us) of an actual random process whose outcome  displays the expected scaling properties in correlation functions. Although it exhibits such a non-trivial structure, it is yet simple enough to allow for a wide range of explicit predictions. 

Utilizing the mean field approximation specifically enables the derivation of closed analytical forms for crucial parameters, such as Euler number densities and their correlations across significant separations: equations~\eqref{eq:EulerNDflight} and~\eqref{eq:bresults}, along with corresponding quadratures for extreme value counts, represent the primary outcomes of this study.
Importantly, our predictions hold irrespective of the density field's variance. While aligning with expectations for a Gaussian field in the regime of small variances, our numerical experiments confirm their validity across all variance values.

Several key insights emerge from our findings:
\begin{enumerate}
\item    Critical point density depends\footnote{They also depend on the specificities of the model through 
structural quantities encoded for instance in the scale cumulant generating function in the context of the large-deviation principle, \citep[see e.g.][]{2016PhRvD..94f3520B}.} on three quantities related to density variance, gradient variance, and variance of the second-order derivative, denoted as $\xi_0$, $\xi_1$, and $\xi_2$, respectively. 
These can be re-expressed using two dimensionless parameters, $\xi_0$ and $\gamma=\xi_1/\sqrt{\xi_0 \xi_2}$, along with $\xi_1$, inversely proportional to distance squared. This implies that number densities scale as $\xi_1^{N/2}$ in N dimensions. Remarkably, while extrema number counts correlate with both $\xi_0$ and $\gamma$, Euler number counts remain independent of $\gamma$. This hints at the impact of small-scale features on peak counts, creating pairs of peaks and troughs at roughly the same height, ultimately leaving Euler number densities unaffected. We speculate that this property may extend to all hierarchical models.
\item    Concerning point correlations, akin to matter fields, we find that critical points of all types exhibit a factorized structure in the large separation limit, as given in \eqref{eq:n1n2fact}. This pattern resembles the separate universe approach and has been validated through our numerical experiments. It enables the derivation of bias parameters for various types of points, as given in particular in \eqref{eq:bresults}.
\item    Notably, bias parameters of maxima, Euler points, and thresholded density fields converge to the same limit in the high-density limit. This suggests that regions with rare high-density values likely host one extremum, sharing identical correlation properties in this limit. This supports simplified approaches where peak correlations mirror those of high-density regions in more complex systems \citep[see for instance][]{1999A&A...349..697B}.
\end{enumerate}

Beyond the paper's scope, extending equations~\eqref{eq:EulerNDflight},\eqref{P1Dextrema} and~\eqref{eq:bresults} to higher dimensions, other Minkowski functionals,  and cross-validating them via Monte Carlo techniques, as done in 1D, will prove valuable
and will be the topic of upcoming papers. One interesting line of investigation is for instance to use these results as benchmarks for validating quasi-Gaussian expansion schemes. It could also be contrasted with those derived in the large deviation limit \citep[as presented in ][]{10.1093/mnras/stw3221}. Furthermore, these statistics may find application in analyzing the influence of Rayleigh-Levy flights in anomalous diffusive processes \citep[e.g. in supersonic turbulence,][]{10.1093/mnras/stx261}  or finite star effects in gravitating systems \citep{Chavanis2009}.

Given their common occurrence, the comprehensive findings from this study hold promise not just in astrophysics but also in broader applications beyond the field.

\section*{Acknowledgements}

We warmly thank S. Appleby for feedback and numerical checks in 2 and 3D. 
We also thank D. Pogosyan for comments and for updating his extrema count codes to suit our purpose, T. Abel for interesting feedback 
and Patrick Peter for  typographic advice. 
We are grateful for the KITP for hosting the workshop
\href{https://www.cosmicweb23.org}{{CosmicWeb23}} during which this project was advanced.
This work is partially supported  by the National Science Foundation under Grant No. NSF PHY-1748958.
This work has made use of the Horizon cluster hosted by the Institut d'Astrophysique de Paris. We also thank S.~Rouberol for  running it smoothly.
 
\section*{Data Availability}
The data underlying this article will be shared on reasonable request to the corresponding author.
Some of the codes underpinning this paper are available on Github at the following URL: \href{https://github.com/cncpichon/levyflight}{https://github.com/cncpichon/levyflight}.


\bibliographystyle{aa}
\bibliography{main} %

\appendix

\section{The Rayleigh Levy flight model}
\label{sec:model}
The random  \flights model was introduced first (in a cosmological context) by J. Peebles (Peebles 1980).  This is a Markov random walk where the PDF of the step length $\ell$ follows a cumulative cumulative distribution function given by
equation~\eqref{eq:defPell},
where $\ell_{0}$ is a small-scale regularization parameter. 
As a result, the density of the subsequent point (first descendant) at position $\vr$ 
is given by
\begin{subequations}
    \begin{align}
f_{1}(\vr)&=\frac{\alpha}{2}\frac{\ell_{0}^{\alpha}}{\vert\vr-\vr_{0}\vert^{1+\alpha}}\ \ \hbox{in 1D space};\\
f_{1}(\vr)&=\frac{\alpha}{2\pi}\frac{\ell_{0}^{\alpha}}{\vert\vr-\vr_{0}\vert^{2+\alpha}}\ \ \hbox{in 2D space};\\
f_{1}(\vr)&=\frac{\alpha}{4\pi}\frac{\ell_{0}^{\alpha}}{\vert\vr-\vr_{0}\vert^{3+\alpha}}\ \ \hbox{in 3D space}.
\end{align}
\end{subequations}

Defining $\psi(k)$ as the Fourier transform of $f_{1}(\vr)$ with equation~\eqref{eq:defpsi}
one can easily show that, assuming there are an infinity of points in the flight, the density in $\vr$ 
of descendants of the point at position  $\vr_{0}$ is given by a series of convolutions whose resummation reads
\begin{equation}
f(\vr_{0},\vr)=\int\frac{\dd^{D}\vk}{(2\pi)^{D}}\ e^{\ii\vk.(\vr-\vr_{0})}\,\frac{1}{1-\psi(k)}.
\end{equation}

The two-point density correlation function is then given by two possible configurations: a neighbourg can either be an ascendant or a descendant, so that the two-point correlation functions between positions $\vr_{1}$ and $\vr_{2}$ are given by 
\begin{align}
\xi_{2}(\vr_{1},\vr_{2})&=\frac{1}{n}\left[f(\vr_{1},\vr_{2})+f(\vr_{2},\vr_{1})\right]
\end{align}
where $n$ is the number density of points in the sample which leads to eq.~\eqref{exactxiRL} of the main text. We note that the two-point correlation within the sample therefore scales
like $1/n$. The latter can be associated with a typical length $\ell_{n}$,
\begin{equation}
n=\frac{1}{\ell_{n}^{D}}.
\end{equation}
At large distance (compared to $\ell_0$), we expect the two-point correlation to behave as power laws. They are given by 
\begin{subequations}
    \begin{align}
\xi_{2}^\mathrm{1D}(r)&=\frac{2 \tan \left(\frac{\pi  \alpha }{2}\right)}{\pi }r^{\alpha -1} \ell_{0}^{-\alpha }\ell_{n}\,,\label{axiRL1D}\\
\xi_{2}^\mathrm{2D}(r)&=\frac{\alpha}{\pi }\ r^{\alpha -2} \ell_{0}^{-\alpha }\ell_{n}^{2}\,,\label{axiRL2D}\\
\xi_{2}^\mathrm{3D}(r)&=\frac{\left(1-\alpha ^2\right) \tan \left(\frac{\pi  \alpha }{2}\right)}{\pi ^2}\ r^{\alpha -3} \ell_{0}^{-\alpha}\ell_{n}^{3}
\label{axiRL3D}.
\end{align}
\end{subequations}
For practical purposes, we give here the resulting expression of the average correlation, $\xi_0$,  
for a Gaussian window function of width $r_s$, 
\begin{subequations}
\begin{align}
\xib_{G}^\mathrm{1D}(r_s)\!\!&=\!\frac{2 \sin \left(\frac{\pi  \alpha }{2}\right) \Gamma \left(\frac{1}{2}-\frac{\alpha }{2}\right) \Gamma
   (\alpha ) r_0^{-\alpha } r_s^{\alpha -1} \ell_{n}}{\pi ^2}\,,\label{axib1D}\\
\xib_{G}^\mathrm{2D}(r_s)\!\!&=\!\frac{2^{\alpha -2} \alpha  \Gamma \left(\frac{\alpha }{2}\right) r_0^{-\alpha } r_s^{\alpha -2}\ell_{n}^2}{\pi }\,,\label{axib2D}\\
\xib_{G}^\mathrm{3D}(r_s)\!\! &=\!\frac{(\alpha +1) \sin \left(\frac{\pi  \alpha }{2}\right) \Gamma \left(\frac{3}{2}-\frac{\alpha
   }{2}\right) \Gamma (\alpha ) r_0^{-\alpha } r_s^{\alpha -3}\ell_{n}^3}{\pi ^3}\label{axib3D}\,.
\end{align}
\end{subequations}
The main interest of this model however lies in the fact that its higher order correlation functions can also be computed and that
they take a simple form. The reason is that $n$ points are correlated when they are embedded in a chronological sequence (that can be run in one direction or the other). Thus the three-point function is simply given by 
\begin{equation}
\xi_{3}(\vr_{1},\vr_{2},\vr_{3})=\frac{1}{n^{2}}\left[f(\vr_{1},\vr_{2})f(\vr_{2},\vr_{3})+...\right]
\label{xi3expression1}
,\end{equation}
with five other terms obtained by all permutations of the indices. Expressing the result in terms of the two-point function, we have 
\begin{align}
\xi_{3}(\vr_{1},\vr_{2},\vr_{3})&=\frac{1}{2}\left[\xi_{2}(\vr_{1},\vr_{2})\xi_{2}(\vr_{2},\vr_{3})+\right.\nonumber\\
&\hskip -1cm\left.\xi_{2}(\vr_{2},\vr_{3})\xi_{2}(\vr_{3},\vr_{1})+\xi_{2}(\vr_{3},\vr_{1})\xi_{2}(\vr_{1},\vr_{2})
\right]
\label{xi3expression2}
.\end{align}
The type of expansion can be pursued to any order. The resulting shapes of the $p$-point correlation function is the following
\begin{align}
\xi_{p}(&\vr_{1},\cdots\vr_{p})=\frac{1}{n^{p-1}}
\!\!\!\!\!\sum_{\sigma\in {\rm perm}[p]}\!\! f(\vr_{\sigma_{1}},\vr_{\sigma_{2}})\dots f(\vr_{\sigma_{n-1}},\vr_{\sigma_{n}})\,, \notag
\end{align}
where $\sigma$ is any permutation of the $p$ indices $1,\dots,p$. It implies that the p-point correlation function
can be expressed in terms of the 2-point functions
\begin{align}
\xi_{p}(&\vr_{1},\cdots\vr_{p})=\left(\frac{1}{2}\right)^{p-2}
\!\!\!\!\!
\sum_{\sigma\in {\rm perm^*}[p]}\!\!\!\!\!\!\!\xi_{2}(\vr_{\sigma_{1}},\vr_{\sigma_{2}})\dots\xi_{2}(\vr_{\sigma_{n-1}},\vr_{\sigma_{n}}), \nonumber
\end{align}
where the $*$ exponent refers to the subset of permutations that lead to a unique un-oriented sequence   (identifying for instance $(1,2,3)$ and $(3,2,1)$).
As we will see hereafter it corresponds to a specific hierarchical tree model. 

\section{The hierarchical tree models}
\label{sec:HTM}

The Rayleigh-Levy flight model is one representative of a large class of models, the so-called hierarchical tree models, that have been put forward in cosmology as a way to model the density field in the highly nonlinear regime. Such models has been  presented in details in \cite{1999A&A...349..697B}. We recall here how they are defined, together with the basic equation that allows the derivation of their cumulant-generating function. Hierarchical tree models are a general class of non-Gaussian fields whose n-point correlation  functions, $\xi_{p}(\vr_{1},\dots,\vr_{p})$, follow the so-called hierarchical Ansatz
\begin{equation}
\xi_{p}(\vr_{1},\dots,\vr_{p})=\sum_{t\in\trees}Q_{p}(t)\,\prod_{\lines\in t}\xi(\vr_{i},\vr_{j}),
\end{equation}
where $\xi(\vr_{i},\vr_{j})$ is the two-point function, while the sum is made over all possible trees that join the $p$ points (diagram without loops), and the tree value, $Q_p$, is obtained by the product of a fixed weight (that depends only on the tree's topology), and the product of the two-point correlation functions, for all pairs that are connected together in the given tree. More specifically it is assumed that
\begin{equation}
Q(t)=\prod_{\vertices\in t} \nu_{p},
\end{equation}
where $\nu_{p}$ is a weight attributed to all vertices with $p$ incoming lines (assuming $\nu_{0}=\nu_{1}=1$ for completion).
In this formalism, the vertex generating function is generally introduced as
\begin{equation}
\zeta(\tau)=\sum_{p}\frac{\nu_{p}}{p!}\tau^{p}. \label{eq:defzeta}
\end{equation}
Such models are thus entirely defined by i) the two-point functions $\xi(r)$ and ii) the vertex-generating function
$\zeta(\tau)$. What the previous section has shown is that the Rayleigh-Levy flight model (in the continuous limit)  
correspond to a specific hierarchical tree model, with $\nu_{2}=1/2$, $\nu_{q}=0$ for $q\ge 3$.

\subsection{Expression for  Cumulant Generating Function }

The exact generating function of multiple cell correlation functions can be built through simple transforms. 
We hereafter consider a set of $n$  cells $i$ of profile $\mW_{i}(\vx)$ (allowing therefore for generic types of profiles, and not only top-hat boxes as was  assumed in early papers). These profiles can obviously overlap but they are assumed to be well localized.  The joint cumulants we consider in this formalism are those of the average densities in cells ${i}$ that can be expressed in terms of spatial averages of correlation functions
\begin{align}
\langle\rho_{1}^{p_{1}}\dots\rho_{n}^{p_{n}}\rangle_{c}&=
\int{\dd\vx_{1,1}}\,\mW_{1}(\vx_{1,1})\dots
\int{\dd\vx_{1,p_{1}}}\,\mW_1(\vx_{1,p_{1}})
\dots
\nonumber\\
&
\dots
\int{\dd\vx_{n,1}}\,\mW_{n}(\vx_{n,1})\dots
\int{\dd\vx_{n,p_{n}}}\,\mW_n(\vx_{n,p_{n}})
\nonumber\\
&\times\xi_{p}(\vx_{1,1},\dots,\vx_{1,p_{1}},\dots\vx_{n,1},\dots,\vx_{n,p_{n}})
.\end{align}
We then wish to build the cumulant-generating function (CGF) of the densities in each cells. It is given by
\begin{align}
\varphi(\lambda_{1},\dots,\lambda_{n})&=
\log
\langle
e^{\sum\lambda_i\rho_i}\rangle \,,\label{CGFdef1}\\
&= 
\sum_{p_{1},\dots,p_{n}}\langle\rho_{1}^{p_{1}}\dots\rho_{n}^{p_{n}}\rangle_{c}\,\frac{\lambda_{1}^{p_{1}}}{p_{1}!}
\dots\frac{\lambda_{n}^{p_{n}}}{p_{n}!}.\label{CGFdef2}
\end{align}
This function represents the generating function of (averaged) cumulants, where the power of the  $\lambda_{i}$ counts the number of points in each cells.  
When the cumulants follow the tree structure described in the previous paragraph, each term that appear in this function then corresponds to   a specific tree. Following a method pioneered by \cite{1987JdC}, \cite{1992A&A...255....1B} showed that it is actually possible to perform the summation over such a set of trees, leading to a formal expression we re-derive briefly below. 

To do so, let us first now define a larger class of objects, $\phi[\nu_p(\vx),\xi(\vx,\vx')]$, (note the different symbol) which represents the sum of trees with vertices of order $p$ having the weight $\nu_p(\vx)$ and lines having the weight $\xi(\vx,\vx')$, and where $\vx$ are space variables that are subsequently integrated in the whole domain. In the context of tree theory, a point that reaches a one-point vertex $\nu_1(\vx)$ are usually called a "leaf". The function we are interested in, $\varphi(\lambda_{1},\dots,\lambda_{n})$, is precisely equal to $\phi[\nu_p(\vx),\xi(\vx,\vx')]$  for the specific choice
\begin{equation}
    \nu_p(\vx)=\nu_p\sum_i\lambda_i\mW_i(\vx)\,.
    \label{eq:nupdex}
\end{equation}

Let us now consider $\phi[\nu_p(\vx),\xi(\vx,\vx')]$ as a functional of $\nu_1(x)$ only, the other quantities it depends on being fixed. 
We can then define $\tau(\vx)$ as the functional derivative of $\phi$ with respect to $\nu_1(\vx)$
\begin{equation}
    \tau(\vx)=\frac{\delta\phi}{\delta\nu_1(\vx)}\,.\label{taudef}
\end{equation}
Then $\tau(\vx)$  appears to be the generating function of all trees with (at least) one leaf at position $x$. This is a sub-part of rooted-trees\footnote{In graph theory rooted trees are trees that emerge from a vertex that have been singled out, not necessarily a leaf.}, and we therefore call $\tau(\vx)$ the rooted-Cumulant Generating Function (r-CGF).
The key relevant property is then that $\tau(\vx)$ obeys a consistency relation, as it can be built recursively: when a line emerges from position $\vx$, it should reach a vertex of order $p\ge 1$ at position $\vx'$ which is then connected to $p-1$ r-CGFs at position $\vx'$. The mathematical transcription of the property is that
\begin{align}
\hskip -0.25cm\tau(\vx)\!=&\!\!\int\!\!\dd\vx'\xi(\vx,\vx')\nu_1(\vx')\!+\! 
   \!\!\int\!\!\!\dd\vx'\xi(\vx,\vx')\!\!\sum_{p=2}^{\infty}\!\!\frac{\nu_p(\vx')}{(p\!-\!1)!}\tau^{p-1}(\vx').
  \!\!\!  \label{tauequation}
\end{align}
Once $\tau(\vx)$ is known,  solving for $\phi$  then simply involves integrating  equation~(\ref{taudef}). This can be done with the help of its Legendre transform with respect to $\nu_1(x)$. 
Defining $\psi$ as  a functional of $\tau(\vx)$ via
\begin{equation}
    \psi=\int\dd\vx\ \tau(\vx)\nu_1(\vx)-\phi, \label{eq:defpsiB}
\end{equation}
the relation (\ref{taudef}) is satisfied when
\begin{equation}
    \frac{\delta\psi[\tau(\vx)]}{\delta\tau(\vx)}
    =\nu_1(\vx)\,.\label{psirelation}
\end{equation}
  As $\xi(\vx,\vx')$  is  a definite positive  operator, the relation (\ref{tauequation}) can be inverted for $\nu_1(\vx)$ as
\begin{equation}
    \nu_1(\vx)=
    \int\dd\vx'\xi^{-1}(\vx,\vx')\tau_1(\vx')\!-\!\sum_{p=2}^{\infty}\frac{\nu_p(\vx)}{(p-1)!}\tau^{p-1}(\vx)\,,
\end{equation}
and formally integrated via equation~\eqref{psirelation} into
\begin{equation}
   \psi=\frac{1}{2}\int\dd\vx\,\dd\vx'\ \tau(\vx)\xi^{-1}(\vx,\vx')\tau(\vx')-\int\dd\vx
    \sum_{p=2}^{\infty}\frac{\nu_p(\vx)}{p!}\tau^{p}(\vx)\,, \notag
\end{equation}
with the correct boundary conditions (it should vanish when $\nu_1(\vx)$ does). From equation~\eqref{eq:defpsiB}, the function $\phi$ then reads 
\begin{equation}
  \hskip -0.1cm \phi\!=\!\int\!\!\dd\vx\sum_{p=1}^{\infty}\!\!\frac{\nu_p(\vx)}{p!}\tau(\vx)^{p}\!-\!\frac{1}{2}\!\int\!\!\dd\vx\dd\vx' \tau(\vx)\xi^{-1}(\vx,\vx')\tau(\vx'),
  \label{eq:PhiEq}
\end{equation}
which solves the formal calculation. 
We can then apply this result for our specific setting (using equation~\ref{eq:nupdex})
to derive the CGF we are interested in.
It is then convenient to re-express $\tau(\vx)$
in  equation~\eqref{eq:PhiEq} in terms of the $\zeta$ function (equation~\ref{eq:defzeta}), the  profiles,  $\mW_{i}(\vx)$  and the  $\lambda_i$ variables. It eventually yields the following expression
\begin{align}
\varphi(\lambda_{1},\dots,\lambda_{n})
&=\sum_{j}\lambda_{j}
\int{\dd\vx}\,\mW_{j}(\vx)\zeta(\tau(\vx))
\nonumber\\
&
-\frac{1}{2}\sum_{j}\lambda_{j}\int{\dd\vx}\,\mW_{j}(\vx)\tau(\vx)\zeta'(\tau(\vx))\,,
\label{ApdC:treesumphi}
\end{align}
and the equation for the r-CGF function reads
\begin{equation}
\tau(\vx)=\sum_{j}\lambda_{j}\int{\dd\vx'}\,\mW_{j}(\vx')\,\xi(\vx,\vx')\,\zeta'(\tau(\vx'))\label{ApdC:treesumtau}.
\end{equation}
The expressions presented in the main text, equations~\eqref{treesumphi} and~\eqref{treesumtau}, are obtained for the specific expression of $\zeta$ corresponding to the Rayleigh-Levy flight statistics, namely $\zeta(\tau)=(1+\tau/2)^2$.
The generic expression for the multiple CGF of hierarchical tree models was presented in \cite{1999A&A...349..697B}.

\subsection{The mean field equation}

To make the resolution of the implicit equations~\eqref{treesumtau} more tractable,
it is possible to assume
that within each cell, $\tau(\vx)$ can be approximated by a constant equal to $\tau_i$.
The consistency equations for the $\tau_i$s, equation~\eqref{treesumtau} then read
 \begin{equation}
\tau_{i}=\sum_{j}\lambda_{j}\xi_{ij}\zeta'(\tau_{j})\,, \label{eq:deftaumeanfield}
\end{equation}
where $\xi_{ij}$ is the average density correlation between cells $i$ and $j$,
\begin{equation}
\xi_{ij}=\int{\dd\vx}\,\mW_{i}(\vx)\ \dd\vx'\,\mW_{j}(\vx')\,\xi(\vx,\vx').
\end{equation}
From equation~\eqref{treesumphi}, we then have
\begin{equation}
\varphi_{n}(\{\lambda_{i}\})=
\sum_{i}\lambda_{i}\left(\zeta(\tau_{i})-\frac{1}{2}\tau_{i}\zeta'(\tau_{i})\right). \label{eq:defvarphimeanfield}
\end{equation}
This is the form we will mostly use in the main text.
\section{The minimal tree model}
\label{sec:MTM}

The minimal tree model (MTM) is defined as a tree model for which only vertices with 2 outcoming lines exist.
It is therefore associated with a vertex generating function of the form
\begin{equation}
\zeta(\tau)=1+\tau+\frac{\nu_{2}}{2}\tau^{2}\,. \label{eq:defzetaRL}
\end{equation}
It has been shown \cite[in][from the behaviour of the void probability function]{1989A&A...220....1B}  that the only possible
value for $\nu_{2}$ is $\nu_{2}=1/2$. This is precisely the case of the Rayleigh Levy flight model.
From equation~\eqref{eq:defzetaRL}, this model is thus characterized by
\begin{equation}
\zeta_{\MTM}(\tau)=\left(1+\frac{\tau}{2}\right)^{2}.
\end{equation}
In this case the stationary equations for a set of cells, equation~\eqref{eq:deftaumeanfield} reads
\begin{equation}
\tau_{i}=\sum_{j}\lambda_{j}\xi_{ij}(1+\tau_{j}/2)\,,\label{tauMTM}
\end{equation}
and equation~\eqref{eq:defvarphimeanfield} becomes
\begin{equation}
\varphi_{n}(\{\lambda_{i}\})=\sum_{i}\lambda_{i}\left(1+\frac{\tau_{i}}{2}\right). \label{eq:defphinMTM}
\end{equation}
As the right hand side of  equation (\ref{tauMTM}) is linear in $\tau$, this system can be solved by the simple inversion
of a matrix. In practice it is therefore relatively easy to derive the expression of $\tau_{i}$
(and therefore of the cumulant generating function) for a finite number of cells.

\subsection{The mean field solution}

From equations~\eqref{tauMTM}-\eqref{eq:defphinMTM} when only one cell is considered,
the one point mean field solution turns out to be
\begin{equation}
\varphi_{1}(\lambda_{\rho};\xi_{0})=\frac{2 \lambda _{\rho }}{2-\xi _0 \lambda _{\rho }}\,.
\label{eq:varphi1}
\end{equation}
It gives the expression of the CGF for a single variable $\rho$ of mean unity and mean square $\xi_{0}$. 
The successive reduced cumulants of such a quantity can be easily computed by Taylor expansion.

Note that
if the random variable $\rho$ is rescaled to have a mean of $\rhob$ (instead of one), its CGF is
$\varphi_{1}(\rhob\lambda;\xi_{0})$, which, given equation~\eqref{eq:varphi1}, is equal to $\rhob\varphi_{1}(\lambda;\rhob\xi_{0})$.

\subsection{MTM composition law and scale convolution}

Let us define generically the cumulant generating function $\varphi_{n}(\{\lambda_{i}\};\{\xi_{ij}\})$ of a set of $n$ cells, of variables $\lambda_{i}$ for cells whose 2-point correlations are $\xi_{ij}$. Then the following identity is satisfied by $\varphi_{n}$
\begin{equation}
\varphi_{n}(\{\lambda_{i}\};\,\{\xi_{ij}+\xi\})=\varphi_{1}(\varphi_{n}(\{\lambda_{i}\};\{\xi_{ij}\});\, \xi).
\label{scalecomposition}
\end{equation}
This important scaling, that we will use repetitively throughout, may seem  at first view as an awkward relation, as it states that the $\xi$ in $\varphi_{n}$ can be shifted by successive application of the function $\varphi_{1}$. This  result seems specific to  \flights.

Let us first briefly demonstrate the property. To avoid confusion in the notation let us define $\tau_{i}$
and $\hat\tau_{i}$ as the solutions of the following systems
\begin{subequations}
    \begin{align}
\tau_{i}&=\sum_{j}\lambda_{j}\xi_{ij}(1+\tau_{j}/2)\,,\\
\hat\tau_{i}&=\sum_{j}\lambda_{j}(\xi_{ij}+\xi)(1+\hat\tau_{j}/2)\,.
\end{align}
\end{subequations}
Then let us also define $\mu_{i}$ and $\hat\mu_{i}$ as
$\mu_{i}=1+{\tau_{i}}/{2}$ and $\hat\mu_{i}=1+{\hat\tau_{i}}/{2}$
so that
\begin{subequations}
\begin{align}
\varphi_{n}&\equiv\varphi_{n}(\{\lambda_{i}\};\{\xi_{ij}\})=\sum_{i}\lambda_{i}\mu_{i}\,,\\
\hat\varphi_{n}&\equiv\varphi_{n}(\{\lambda_{i}\};\{\xi_{ij}+\xi\})=\sum_{i}\lambda_{i}\hat\mu_{i}.
\end{align}
\end{subequations}
Then $\hat\mu_{j}$ is the only solution of the system
\begin{equation}
\sum_j M_{ij}\hat\mu_{j}=1\label{MijEqs}
\,,
\end{equation}
where the matrix $M_{ij}$ is defined as
\begin{equation}
M_{ij}=\left(\delta_{ij}-\frac{1}{2}\sum_{j}\lambda_{j}(\xi_{ij}+\xi)\right).
\end{equation}
Now, since
\begin{equation}
\sum_j  M_{ij}\mu_{j}=1-\frac{\xi}{2}\sum_{j}\lambda_{j}\mu_{j}\,,
\end{equation}
this implies that ${\mu_{i}}/\left({1-\frac{\xi}{2}\sum_{j}\lambda_{j}\mu_{j}}\right)$
is the solution of equation~\eqref{MijEqs}, so that
by identification we have
\begin{equation}
\hat\mu_{i}=\frac{\mu_{i}}{1-\frac{\xi}{2}\sum_{j}\lambda_{j}\mu_{j}}.
\end{equation}
Hence, 
\begin{equation}
\hat\varphi_{n}=\sum_{i} \lambda_{i}\frac{\mu_{i}}{1-\frac{\displaystyle \xi}{2}\sum_{j}\lambda_{j}\mu_{j}}
=\frac{\varphi_{n}}{1-\frac{\xi}{2}\varphi_{n}}=\varphi_{1}(\varphi_{n};\xi)\,,
\end{equation}
which establishes the relation.

Equation~\eqref{scalecomposition} reflects some interesting  physical properties it is related to: scale composition when cell correlations are built as a two-steps procedure. Indeed, let us consider a set of random \flights experiments, in which instead of having of fixed number of points in each sample, the sample density is itself drawn from the one-point PDF derived from an MTM process. We denote $\rho_{s}$ the sample density (whose average is set to unity). Its CGF is then $\varphi_{1}(\lambda_{s};\xi_{s})$,  where $\xi_{s}$ is the variance of the sample density.

We then note that for each realization the cell densities, $\rho_{i}$ scale like $1/\rho_{s}$ (by definition $\rho_{i}$ is defined with respect to the mean of the survey) and that its correlation functions $\xi_{ij}$ scale like $1/\rho_{s}$ (this is a consequence of the expression of the scaling of the two-point function in the Rayleigh Levy model). 
To be more precise let us define $\xib_{ij}$ as the cell correlations when the sample density is equal to unity. We then have
\begin{equation}
\xi_{ij}=\xib_{ij}/\rho_{s}.
\end{equation}
Let us then define the densities $\hat\rho_{i}$ which represent the ``true'' density in the sample (that is when the sample density is taken into account) as
\begin{equation}
\hat\rho_{i}=\rho_{s}\,\rho_{i},
\end{equation}
and aim at building the joint PDF of $\hat\rho_{i}$. This PDF is formally given by
\begin{align}
P&(\{\hat\rho_{i}\})=\int\frac{\dd\rho_{s}}{\rho_{s}^{n}}P(\rho_{s})
P(\{\rho_{i}/\rho_{s}\})\,,\\
&=
\int\frac{\dd\rho_{s}}{\rho_{s}^{n}}P(\rho_{s})
\!\!\int\!\!\Pi_{i}\frac{\dd\lambda_{i}}{2 \pi \ii} 
\exp\left(-\sum_{i}\lambda_{i}\hat\rho_{i}/\rho_{s}+
\varphi_{n}(\{\lambda_{i}\};\{\xi_{ij}\})\right)\,, \notag
\end{align}
where the dependence in $\rho_{s}$ is also present in the expression of $\xi_{ij}$.
Then, making the change of variable
\begin{equation}
\hat\lambda_{i}=\lambda_{i}/\rho_{s}\,,
\end{equation}
and  noting that 
\begin{equation}
\varphi_{n}(\{\lambda_{i}\};\{\xi_{ij}\})=\rho_{s}
\varphi_{n}(\{\hat\lambda_{i}\};\{\xib_{ij}\})\,,
\end{equation}
we have
\begin{align}
P(\{\hat\rho_{i}\})&=\int{\dd\rho_{s}}\int\frac{\dd\mu}{2 \pi \ii}\int\Pi_{i}\frac{\dd\lambda_{i}}{2 \pi \ii}\nonumber\\
&\hskip -1.cm\times\exp\left(-\sum_{i}\hat\lambda_{i}\hat\rho_{i}
-\mu\rho_{s}+
\rho_{s}\varphi_{n}(\{\hat\lambda_{i}\};\{\xib_{ij}\})
+\varphi_{1}(\mu,\xi_{s})\right).
\end{align}
Now the integral over $\rho_{s}$ leads to a Dirac delta function in $\mu-\varphi_{n}(\{\hat\lambda_{i}\};\{\xib_{ij}\})$ leading to
\begin{eqnarray}
P(\{\hat\rho_{i}\})\!=\!\!\int\!\!\Pi_{i}\frac{\dd\lambda_{i}}{2 \pi \ii}
\exp\left(\!-\!\!\sum_{i}\!\hat\lambda_{i}\hat\rho_{i}
\!+\!\varphi_{1}(\varphi_{n}(\{\hat\lambda_{i}\};\{\xib_{ij}\}),\xi_{s})\!\right)\!. \label{eq:formaldefPDF}
\end{eqnarray}
This expression means formally that the CGF of $\hat\rho_{i}$ is this two-step construction given by  $\varphi_{1}(\varphi_{n}(\{\hat\lambda_{i}\};\{\xib_{ij}\}),\xi_{s})$. The relation (\ref{scalecomposition}) ensures that it is also given by $\varphi_{n}(\{\hat\lambda_{i}\};\{\xib_{ij}+\xi_{s}\})$, which states that survey density fluctuations can, in this model, be taken into account via a simple shift in the cell correlation amplitudes.

A useful practical consequences of property (\ref{scalecomposition}) is that one could set any peculiar element of $\xi_{ij}$
to zero. For instance the CGF of 2 identical cells of density variance $\xib$ and of cross correlation $\xi_{12}$ is given by
\begin{equation}
\varphi_{2}(\lambda_{1},\lambda_{2};\xib,\xi_{12})=
\varphi_{1}(\varphi_{1}(\lambda_{1};\xib-\xi_{12})+\varphi_{1}(\lambda_{2};\xib-\xi_{12});\xi_{12})\,. \notag
\end{equation}
It is possible to extend such a construction for a larger number of cells\footnote{up to 3 cells at 1D, 5 cells at 2D and 7 cells at 3D.}, but for specific configurations only.  Note finally that when one needs to construct the density CGF for a large number of cells it is convenient to set $\xib=0$ for all cells as it makes the system in $\tau_{i}$ sparser.

\subsection{Computing PDF in the minimal tree model}

The basic quantity one wishes to have access to is the PDF, defined as  the  inverse Laplace transform of the cumulant generating function $\varphi_{1}(\lambda_{\rho})$, that is
\begin{equation}
P(\rho)=\int_{-\ii\infty}^{+\ii\infty}\frac{\dd \lambda_{\rho}}{2\pi\ii}\exp\left(-\lambda_{\rho}\,\rho+\varphi_{1}(\lambda_{\rho})\right)\,,
\end{equation}
where the integral runs formally along the imaginary axis, but can be moved in the complex plane as long as no poles or singularities are encountered along the path. 

The following  relation  will be exploited throughout this paper
\begin{equation}
\hskip -0.1cm \!\int\!\frac{\dd \lambda_{\rho}}{2\pi\ii}
\exp\left(\!{-\lambda_{\rho}\,\rho\!+\!\frac{a_{c}}{\lambda_{c}\!-\!\lambda_{\rho}}}\!\right)
\!=\!\Dirac(\rho)\!+\! a_{c}e^{-\lambda_{c}\rho}
\, \tHyp\!\left(2;a_{c} \rho\right),
\label{KeyLapInv}
\end{equation}
where $\Dirac$ is the one-dimensional Dirac distribution and $\tHyp(a;z)$ is the regularized confluent hypergeometric function 
$\,_{0}F_{1}(a;z)/\Gamma(a)$.  This formula can be applied directly to $\varphi_{1}(\lambda_{\rho};\xi_{0})$ to derive $P(\rho)$. It can also be applied after successive derivatives with respect to $a_{c}$ from which on can establish that
\begin{align}
\int\frac{\dd \lambda_{\rho}}{2\pi\ii}\left(\frac{1}{\lambda_{c}-\lambda_{\rho}}\right)^{n}
&\exp\left(-\lambda_{\rho}\,\rho+\frac{a_{c}}{\lambda_{c}-\lambda_{\rho}}\right)
=\nonumber\\
&
e^{-\lambda_{c}\rho}\left(\frac{\dd}{\dd \,a_{c}}\right)^{n}\left\{a_{c}\, \tHyp\left(2;a_{c} \rho\right)\right\}\,\label{KeyLapInv2},
\end{align}
for $n\ge 1$.
Then, the application of equation~(\ref{KeyLapInv}) to the expression (\ref{eq:varphi1}) leads to
\begin{equation}
P(\rho)=e^{-2/\xi _0}\Dirac(\rho)+\frac{4} {\xi _0^2}e^{-\frac{2 (\rho +1)}{\xi _0}} \, \tHyp\left(2;\frac{4 \rho }{\xi _0^2}\right)\,,
\end{equation}
which yields the PDF in the mean field approximation. One can see that it involves the sum of two terms, a Dirac term term for $\rho=0$, and a continuous contribution. This highlights one of the key feature of this model, which  is that  the probability of having empty regions of finite size remains finite even in the continuous limit.

\subsection{Beyond the mean field approximation}
Let us explore the possibility of solving the consistency relations beyond the mean field approximation, which provides means to explore the validity of this approximation. The calculations will be limited to the 1D case and to a Gaussian filter and its derivatives.  We therefore assume that 
\begin{equation}
    \mW_{0}(x)=\frac{1}{\sqrt{2 \pi}}
    \exp\left(-\frac{x^2}{2}\right).
\end{equation}
The idea is then to expand $\tau(x)$ in e.g. equation~\eqref{treesumtau} on its natural ortho-normal basis, namely the basis of the Hermite Polynomials.
More precisely, $\tau(x)\mW_{0}(x)$ being bounded at large distance, it is possible to expand it as 
\begin{equation}
    \tau(x)=\sum_{n=0}^{\infty}
    \tau_n\,{h_n(x)}, 
    \,\,\mathrm{with }
\,\,
    h_n(x)=\frac{1}{2^{n/2}\sqrt{n!}} H_n\left(\frac{x}{\sqrt{2}}\right),
\end{equation}
noting that 
\begin{equation}
    \int\dd x\,
    {h_n(x)}\,
    {h_{n'}(x)}\,\mW_{0}(x)=\delta_{nn'}.
\end{equation}
We are  specifically interested in the joint CGF, $\varphi(\{\lambda_0,\lambda_1,\lambda_2\})$ involving fields dual to the density, the first and second derivatives. The window functions for the latter  can be expressed in terms of the Hermite polynomials  as
\begin{equation}
    \mW_{q}(x)=\left(\frac{\dd}{\dd x}\right)^q\mW_{0}(x)=\sqrt{q!}\,h_q(x)\mW_{0}(x)\,,
\end{equation}
for $q=1$ and $q=2$, and we have
\begin{equation}
    \int\dd x\, \mW_{q}(x)
    {h_n(x)}=\sqrt{q!}\delta_{qn}.
\end{equation}
As a result, the consistency relation~\eqref{treesumtau} for $\tau(x)$  now reads
\begin{eqnarray}
\tau(x)\!=\!\int\!\dd x'\xi(x,x')\left(1+\frac{\tau(x')}{2}\right)
\mW_{0}(x')\sum_{q=0}^2\lambda_q \sqrt{q!}\,h_q(x'),
\end{eqnarray}
which can then be transformed into a system in $\tau_p$ after integration with weight $h_p(x)C_0(x)$
\begin{align}
\tau_p=\sum_{q=0}^2\lambda_q
\left[\Xi_{p0}^{(q)}+\sum_{p'=0}^{\infty}\frac{\tau_{p'}}{2}\Xi_{pp'}^{(q)}\right],
\end{align}
with
\begin{equation}
\Xi_{pp'}^{(q)}=\int\dd x\ \dd x'\ 
\mW_{0}(x)
{h_p(x)}
 \xi(x,x')
{h_{p'}(x')}
\mW_{q}(x')\,.
\end{equation}
The expression of CGF for the density and its gradients derives from equation~\eqref{treesumphi}, and is simply given by
\begin{equation}
    \varphi(\{\lambda_q\})=\lambda_0+\sum_{q=0}^2\lambda_q\frac{\tau_q}{2}\sqrt{q!}\,,
\end{equation}
taking advantage of the orthogonality relations between $\mW_{q}(x)$ and $h_p(x)$.
Note that those quantities depend on the actual shape and amplitude of the correlation function. In the following, we simply assume that $\xi(x,x')\sim\vert x-x'\vert^{-1/2}$ corresponding to $\alpha=1/2$ for a 1D Rayleigh-Levy flight. It implies that $\Xi_{pp'}^{(q)}/\xi_0$ are all fixed quantities and that the result can therefore be expressed in term of $\xi_0$ only. 

We present hereafter the result of such derivations for the density,
when the Hermite expansion in $\tau$ is truncated at increasing order, from 0 (where the r-CGF is assumed to be constant) to 4 
(in practice we have been able to derive the CGF up to 10th order as illustrated in the figures below). We thus have
\begin{subequations}
\begin{align}
\hskip -0.2cm
\varphi_0(\lambda)&\!= \frac{2 \lambda }{2-\lambda  \xi _0}\,,\\
\varphi_2(\lambda)&\!=   -\frac{\lambda  \left(5 \lambda  \xi _0-64\right)}{2 \lambda ^2 \xi _0^2-37 \lambda  \xi _0+64}\,,\\
\varphi_4(\lambda)&\!=\! 
-\frac{3 \lambda  \left(75 \lambda ^2 \xi _0^2\!-\!8240 \lambda  \xi _0+65536\right)}{80 \lambda ^3 \xi _0^3\!-\!10849 \lambda ^2 \xi
   _0^2+123024 \lambda  \xi _0-196608},
\end{align} \label{eq:beyond}
\end{subequations}
for subsequent truncation orders. The first expression reproduces the one-cell mean field approximation. The others correspond to corrections to it. 

\begin{figure}
\includegraphics[width=0.99\columnwidth]{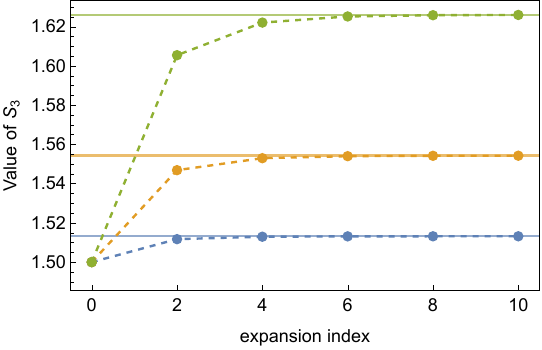}
\caption{Value of the reduced skewness, $S_3$,  obtained at increasing order beyond the mean field. The mean field solution of the \flights model, $3/2$,
is the same, whatever the index of the spectrum and the shape of the window function. The exact expression of $S_3$  depends however slightly on the power law index and on the filter shape. They are indicated as dotted lines. One can see that corrections to the mean field solution converge very rapidly to the expected value.}
\label{fig:S3_MFExpand}
\end{figure}

These expressions exhibit a number of  worthwhile properties. They predict values of reduced cumulants that slightly evolve with the order of the truncation. This is illustrated on Figure \ref{fig:S3_MFExpand} for the skewness. It is actually possible to compute its exact expression  for the Rayleigh-Levy flight model with a given slope for the two-point function -- or equivalently the index of the power spectrum -- and a given filter shape -- here a Gaussian filter. Figure \ref{fig:S3_MFExpand} shows that the prediction from the Mean Field expansion rapidly converges to the exact value. The rapid convergence can also be observed for higher order cumulants. This suggests that the high density tail of the PDF is well captured by Mean Field approach.

\begin{figure}
\includegraphics[width=0.99\columnwidth]{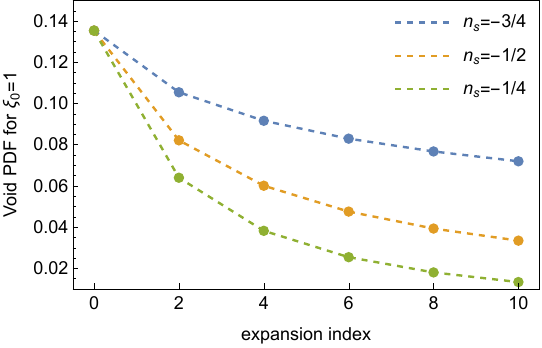}
\caption{Value of the Void Probability Density Function  for $\xi_0=1$ at increasing order beyond mean field solution. The expression of the VPDF depends both on the power law index and on the filter shape. For a Gaussian filter one expects the VPDF to identically vanish, since they have infinite spatial extensions, unlike filters with compact support. It makes the prediction derived from the mean field  poor  in the low density regime.}
\label{fig:VPDF_MFExpand}
\end{figure}

On the other hand the low density part turns out to be much more difficult to capture with a Gaussian filter. As illustrated on Figure \ref{fig:2Drealization}, the   density fields exhibit genuine large empty regions. This leads to a non-zero Void Probability Density Function for filters that have a compact support. However, for filters with extended radial tails, the probability of finding the filtered density to be exactly zero can only vanish, leading however to an excess of probability at low densities. This is illustrated in Figure \ref{fig:PDFMFExpand}. 
Unfortunately the mean field extensions as described here always lead to finite VPDF to all finite orders. This effect is illustrated in Fig.~\ref{fig:VPDF_MFExpand}, which also shows that the convergence of the PDF in the low density regions is slow.
This is also illustrated at the level of the density PDF in Fig.~\ref{fig:PDFMFExpand}. 
To conclude,
i) the mean field solution is very efficient at predicting the high density regions but fails in the low density regions; 
ii) solutions  beyond the mean field, equations~\eqref{eq:beyond}, can account for the behaviour of the PDFs in the low density regions, but the convergence is slow. 
Extension   beyond the mean field in higher dimensions will be the topic of future work.
\section{Extrema and Euler densities}
\label{sec:ExtremaDerivation}

\subsection{The general formalism}

The computation of extrema densities is relies on the knowledge of the joint PDF of the local density, its first and second order 
derivatives \citep{1986ApJ...304...15B}. The latter will be derived from the CGF of those quantities.

The number of relevant variables  depend on the dimension of space. To fix the notation we define $\xi_{0}$, $\xi_{1}$ and $\xi_{2}$ as the variance of respectively the local density, the one-component density gradient and second derivatives
\begin{equation}
\xi_{0}=\mg\rho^{2}\mdd-1;\label{xi0def}\quad
\xi_{1}=\mg\rho_{,x}^{2}\mdd;\quad
\xi_{2}=\mg\rho_{,xx}^{2}\mdd\,,
\end{equation}
assuming $\mg\rho\mdd=1$. We further note the consistency relations which can be derived by integration by part
\begin{align}
\mg\rho\rho_{,xx}\mdd=-\xi_{1};\label{avrrxx} \quad
\mg\rho_{,xx}\rho_{,yy}\mdd=\frac{1}{3}\xi_{2};\quad
\mg\rho_{,xy}^{2}\mdd=\frac{1}{3}\xi_{2}.
\end{align}
We then denote $\lambda_{\rho}$, $\lambda_{i}$, and $\lambda_{ij}$ the conjugate variables to respectively $\rho$, $\rho_{,i}$ and
$\rho_{,ij}$ (keeping in mind that $\rho_{,ij}$ and $\rho_{,ji}$ are identical).

In general the number density of extrema is given by equation~\eqref{eq:defeuler}.
For instance, maxima are obtained when the integral is restricted to the regions where all eigenvalues are negative; minima when all eigenvalues are positive and in 2D, saddle points are obtained when the integral is restricted to the regions where the sign of the two eigenvalues is different. We finally 
note that Euler number densities are obtained with $\chi^{\rm extr.}\left[\rho_{,ij}\right]=1$. Its advantage is that it preserves the analytical structure of the operator making in general its derivation easier. Specifically, the formulae
we obtain for the 1D and 2D cases are
\begin{align}
    n_{\rm 1D}^{\rm Euler}(\rho)&=
\int{\dd\rho_{,xx}}\ \rho_{,xx}
    \,P\left(\rho,\rho_{,x}=0,\rho_{,xx}\right)\,,  \label{1DnEuler}\\
     n_{\rm 2D}^{\rm Euler}(\rho)&=
\int{\dd\rho_{,xx},\dd\rho_{,yy},\dd\rho_{,xy}}\!
\left(\rho_{,xx}\rho_{,yy}-\rho_{,xy}^2\right)
   \,P\left(\rho,\rho_{,i}=0,\rho_{,ij}\right)\,,
   \notag \label{2DnEuler}
\end{align}
and a similar - if more involved expression - for the 3D case. For the Euler number density, taking advantage of its analyticity, it is possible to simply re-express it in terms of the CGF through inverse Laplace transforms. More precisely we have in one D
\begin{align}
    n_{\rm 1D}^{\rm Euler}(\rho)&=
\int\frac{\dd\lambda_{\rho}}{2\pi \ii}
\frac{\dd\lambda_{x}}{2\pi \ii}\ \
\left.\frac{\partial}{\partial\lambda_{xx}}\right._{\vert_{\lambda_{xx}=0}}\!\!\!\!\!
\exp(-\lambda_\rho\rho+\varphi(\lambda_\rho,\lambda_x,\lambda_{xx})) \notag
\end{align}
 It shows that the Euler number density, unlike the extrema density in general, depends only on limited information from the joint CGF.
For the 2D case the calculations are similar but a bit more involved,
\begin{align}
    n_{\rm 2D}^{\rm Euler}(\rho)&=
\int\frac{\dd\lambda_{\rho}}{2\pi \ii}
\frac{\dd\lambda_{x}}{2\pi \ii}
\frac{\dd\lambda_{y}}{2\pi \ii}\ \
\left(\frac{\partial^2}{\partial\lambda_{xx}\partial\lambda_{yy}}-\frac{\partial^2}{\partial\lambda_{xy}^2}\right)_{\vert_{\lambda_{ij}=0}}
\nonumber\\
    &
\times\exp(-\lambda_\rho\rho+\varphi(\lambda_\rho,\lambda_i,\lambda_{ij}))\,,
\end{align}
with a similar construction for the 3D case. Explicit derivation of the Euler number densities for the Rayleigh-Levy flight model are presented hereafter. Let us start with general consideration regarding the symmetry properties of both the CGF and the joint PDFs.

\subsection{Spatial homogeneity \& isotropy}
In the cosmological context, density fields are statistically homogeneous and isotropic. For the hierarchical tree models, it is ensured by the fact that the two-point correlation function $\xi_2(\vx,\vx')$ depends on
$\vert\vx-\vx'\vert$ only. 

A number of properties follows. The first immediate one, for the 1D case, is that the expectation value of any gradient should vanish. That implies in particular that the expectation value of $\rho_{,x}^2+\rho\rho_{,xx}$ vanishes, which implies equation~(\ref{avrrxx}). More generally a consequence of this invariance is that, for any integer $p$ and $q$,
\begin{equation}
    p\mg\rho^{p-1}\rho_{,x}^{q+1}\mdd_c+q
    \mg\rho^{p}\rho_{,x}^{q-1}\rho_{,xx}\mdd_c=0\,,
\end{equation}
which in terms, after summing over $p$ and $q$, implies
\begin{equation}
\lambda_\rho\frac{\partial}{\partial\lambda_x}\varphi(\lambda,\lambda_x)
+\lambda_x
{\frac{\partial}{\partial\lambda_{xx}}\Big\vert}_{{\lambda_{xx}=0}}
\varphi(\lambda,\lambda_x,\lambda_{xx})=0\,.
\end{equation}
An alternative formulation at the level of the PDF reads
\begin{equation}
\hskip -0.3cm    \int\!\!\dd\rho_{\!,xx}\left\{\rho_{\!,x}\frac{\partial P(\rho,\rho_{\!,x},\rho_{\!,xx})}{\partial \rho}+
    \rho_{,xx}\frac{\partial P(\rho,\rho_{\!,x},\rho_{\!,xx})}{\partial \rho_{\!,x}}\right\}=0\,. \label{eq:translation-invariance}
\end{equation}
Integrating the second term of this equation over $\rho_{,x}$ up to $\rho_{,x}=0$ leads to the expression (\ref{1DnEuler}) of $n_{\rm 1D}^{\rm Euler}(\rho)$ so that
the latter can eventually be written in terms of the joint probability of the density and its gradient as
\begin{equation}
    n_{\rm 1D}^{\rm Euler}(\rho)=\frac{\partial}{\partial\rho}\int_{-\infty}^{0}
    \rho_{,x}\dd\rho_{,x}P(\rho,\rho_{,x})\,.
\end{equation}
It shows that the Euler number density does not depend actually on the way $\rho_{,xx}$ is correlated to the density and its gradient. This is not so for the extrema counts. This points to an interesting feature of Euler number density:  it is significantly more insensitive to small scale fluctuations compared to other observables. 

Unfortunately this simplification does not directly extend to heigher dimensions.
There are however a number of consistency relations that derive from statistical invariance under translation, parity change and rotation. They are
\begin{enumerate} 
    \item Translation invariance\footnote{for time translation invariance such processes are called stationary, as explored in \cite{Adler2009}.}: as in 1D, for any direction $i$ one expects
\begin{equation}
\lambda_\rho\frac{\partial}{\partial\lambda_i}\varphi(\lambda,\{\lambda_i\})
+\sum_j\lambda_j
\frac{\partial}{\partial\lambda_{ij}}_{\Big\vert_{\lambda_{ij}=0}}\!\!\!\!\!
\varphi(\lambda,\{\lambda_i\},\{\lambda_{ij}\})=0.
\end{equation}
    \item Parity invariance: For the dimensions above 2 there are other combinations that vanish such that the expectation values of $T_{,x}U_{,y}-T_{,y}U_{,x}$ (as can be shown by successive integrations by parts). For the 2D case this is the transcription of
parity invariance: the expectation of any rotational is expected to vanish.
It implies that 
\begin{equation}
    \lambda_y\left(
    \frac{\partial^2\varphi}{\partial\lambda_{xx}\partial\lambda_{yy}}
   \! -\!\frac{\partial^2\varphi}{\partial\lambda_{xy}^2}
    \right)_{\Big\vert_{\lambda_{ij}=0}}\!\!\!\!\!\!=
    \lambda_\rho\left(
    \frac{\partial^2\varphi}{\partial\lambda_{x}\partial\lambda_{xy}}
    \!-\!\frac{\partial^2\varphi}{\partial\lambda_{y}\partial\lambda_{xx}}
    \right)_{\Big\vert_{\lambda_{ij}=0}} \notag
\end{equation}
and a similar equation after substitution $y\leftrightarrow x$. Note that the left-hand-side of this equation is the operator for the computation of the Hessian in 2D. It can be reduced to first order derivatives in $\lambda_{ij}$. Contributions from joint cumulants between the density and its first and second order derivatives still however contribute to the Euler number density.
    \item Rotation invariance: for a sake a completeness one can also derive the consequences of  rotational  invariance. We do it here for the 2D case: forcing $\lambda_{x}\rho_{x}+\lambda_{y}\rho_{y}+
    \lambda_{xx}\rho_{xx}+
    \lambda_{xy}\rho_{xy}+
    \lambda_{yy}\rho_{yy}$ to be rotation invariant, a rotation of the coordinate system by infinitesimal angle $\theta$ leads to 
    \begin{subequations}
     \begin{align}
        \delta\lambda_x&=-\theta\lambda_y\,,\quad
        \delta\lambda_y=\theta\lambda_x\,,\\
        \delta\lambda_{xx}&=-\theta\lambda_{xy}\,,\quad
        \delta\lambda_{yy}=\theta\lambda_{xy}\,,\\
        \delta\lambda_{xy}&=-2\theta(\lambda_{xx}-\lambda_{xx})\,,
    \end{align}     
    \end{subequations}
so that 
\begin{equation}
    \mR\left[\varphi(\lambda_\rho,\lambda_i,\lambda_{ij})\right]=0\,,
\end{equation}
with \begin{align}
\hskip -0.1cm \mR_\theta&\!=\!
\frac{\lambda_x\partial}{\,\,\,\partial\lambda_y}\!-\!\frac{\lambda_y\partial}{\,\,\,\partial\lambda_x}
\!+\!\left(\frac{\lambda_{xy}\partial}{\,\,\,\partial\lambda_{yy}}\!-\!\frac{\lambda_{xy}\partial}{\,\,\,\partial\lambda_{xx}}\right)
\!-\!2\left(\lambda_{yy}\!-\!\lambda_{xx}
\right)\frac{\lambda_y\partial}{\,\,\,\partial\lambda_{xy}}. \notag
\end{align}
\end{enumerate}

Let us now turn  by exploring  scaling properties of the cumulants.  Hierarchical models are by nature such that
\begin{equation}
    \mg\rho^n\mdd_c\sim\xi_0^{n-1}\,.
\end{equation}
In general cumulants of product of powers of the local density, first and second order derivatives behave like $\xi_0^s \xi_1^t \xi_2^u$ where ${s+t+u=n-1}$, if the total power is $n$. Note that for dimensional reasons, if only the local density and first derivative is involved then
\begin{equation}
    \mg\rho^p\rho_x^q\mdd_c\sim\xi_0^{p+q/2-1}\xi_1^{q/2}\,.
\end{equation}
It is not possible to give rules when both density, first and second order derivatives are present. We can however note that
\begin{equation}
    \mg\rho^p\rho_{,xx}\mdd_c=-p\mg\rho^{p-1}\rho_{,x}^2\mdd_c
    \ \sim\ \xi_0^{p-1}\xi_1\,.
\end{equation}
Finally we \textit{conjecture}\footnote{
Indeed,  geometrically, adding high-frequency low-amplitude noise to a ND field, while creating new 
extrema, will also introduce saddle points  along the new persistent ridges (ascending or descending 1D manifolds) linking them to existing extrema 
\protect\citep{2023arXiv230911558C}. This will not 
 change the $n_\mathrm{ND}^\mathrm{Euler}$ of the field, since the signature of the new paired points only differ by one,
 so their contributions cancel.
This in turn  suggests that $\xi_2$ does not contribute to $n_\mathrm{ND}^\mathrm{Euler}$.
} that the following scaling is expected,
\begin{equation}
    \mg\rho^p\rho_{,i\,}^q\rho_{,ij}\mdd_c
   \ \sim\ \xi_0^{p+q/2-1}\,\xi_1^{1+q/2}\,.
\end{equation}
It is indeed verified in the expressions derived below.

\subsection{Derivation of  join CGF in  mean field limit}

Let us now turn to writing the joint CGF for a given set of cells. The calculation is based on the mean field solution for a finite number of cells assumed to be infinitely close to one another; the derivatives being then obtained via finite differences.
More precisely if two cells are at distance $d$, we then assume that the average correlation between the 2 cells, $\xi(d)$, can be expanded
in powers of $d$. Its expression can be derived using the correlation between the density and its gradient (again via integration by part)
\begin{equation}
    \mg\rho\rho_{,xx}\mdd=-\xi_{1},\quad
    \mg\rho\rho_{,xxxx}\mdd=\xi_{2},
\end{equation}
so that
\begin{equation}
\xi(d)\approx \xi_{0}-\frac{1}{2}\xi_{1}d^{2}+\frac{1}{24}\xi_{2}d^{4}\,,
\end{equation}
where the coefficients $\xi_{i}$ are precisely those defined in equation~\eqref{xi0def}.

The density gradient is then represented by a finite difference $(\rho_{i}-\rho_{j})/d$ is the two cells are at distance $d$.
The conjugate variable associated to the density gradient, $\lambda_{x}$, would then contribute to both $\lambda_{i}$ and $\lambda_{j}$ with weight respectively $1/d$ and $-1/d$. In general we then build the multi-values CGF as a function of such quantities. To make the construction more explicit let us consider the 2D case.
To do so we need at least 6 cells taken from a 3x3 grid. Each $\lambda_{\alpha}$ where $\alpha$ stands for $\rho$ or its derivatives contribute through the following pattern
\begin{subequations}
    \begin{align}
L_{\rho}&=\left(
\begin{array}{ccc}
 0 & 0 & 0 \\
 0 & 1 & 0 \\
 0 & 0 & 0 \\
\end{array}
\right)\,, \quad
L_{x}=\frac{1}{2d}
\left(
\begin{array}{ccc}
 0 & 0 & 0 \\
 -{1}& 0 & {1} \\
 0 & 0 & 0 \\
\end{array}
\right)\,,\\
L_{xx}&\!=\!\frac{1}{d^{2}}
\left(
\begin{array}{ccc}
 0 & 0 & 0 \\
 1 & -2 & 1 \\
 0 & 0 & 0 \\
\end{array}
\right)\,, \,\,
L_{xy}\!=\!\frac{1}{d^{2}}
\left(
\begin{array}{ccc}
 0 & -1 & 1 \\
 0 & 1& -1 \\
 0 & 0 & 0 \\
\end{array}
\right),
\end{align}
\end{subequations}
with a similar definition for $L_y$ and $L_{yy}$.
Defining the 3x3 matrix
\begin{equation}
L_{\lambda}=\lambda_{\rho}L_{\rho}+\lambda_{x}L_{x}+\lambda_{y}L_{y}+\lambda_{xy}L_{xy}+
\lambda_{xx}L_{xx}+\lambda_{yy}L_{yy}\,,
\end{equation}
the stationary equations become
\begin{equation}
\tau_{i}=\sum_{j=1}^{9}\xi_{ij}(d)(L_{\lambda})_{j}\left(1+\tau_{j}/2\right)\,, 
\label{eq:tau9}
\end{equation}
where $(L_{\lambda})_{j}$ is the $j$th element of the matrix $L_{\lambda}$. The CGF then reads
\begin{equation}
\varphi(\lambda_{\rho},\lambda_{i},\lambda_{ij})=
\sum_{i=1}^{9}(L_{\lambda})_{i}\left(1+\tau_{i}/2\right).
\end{equation}
The result is a function of $\lambda_{\alpha}$ and of $d$, $\xi_{0}$, $\xi_{1}$ and $\xi_{2}$
through the $\xi_{ij}$ dependence.
The set of 9 equations~\eqref{eq:tau9} can be solved via {\sc \small Mathematica}, yielding a rather complex expression. 
\subsection{The 1D Euler and extrema counts}
In one 1D, in the small $d$ limit and for ${\xi_0=0}$,  the generating 1D function, 
$\varphi_{\rm 1D}(\lambda_{\rho},\lambda_{\rm x},\lambda_{\rm xx};0,\xi_{1},\xi_{2})$, takes  the 
remarkably simple form
\begin{equation}
    -\frac{2 \left(\xi _1 \lambda _x^2+\lambda _{\rho } \left(2-2 \xi _1 \lambda _{xx}\right)+\xi _2 \lambda
   _{xx}^2 \left(1-\xi _1 \lambda _{xx}\right)\right)}{\left(\xi _1 \lambda _{xx}-1\right)
   \left(\xi _1 \lambda _{xx}+2\right){}^2}\,.
\end{equation}
Taking now advantage of the shifting relation (\ref{scalecomposition}), we can  write for a general $\xi_0$
\begin{align}
&\varphi_{\rm 1D}(\lambda_{\rho},\lambda_{\rm x},\lambda_{\rm xx};\xi_{0},\xi_{1},\xi_{2})=\nonumber\\
& 
\varphi_1\left[-\frac{2 \left(\xi _1 \lambda _x^2\!+\!\lambda _{\rho } \left(2\!-\!2 \xi _1 \lambda _{xx}\right)\!+\!\xi _2 \lambda
   _{xx}^2 \left(1\!-\!\xi _1 \lambda _{xx}\right)\right)}{\left(\xi _1 \lambda _{xx}\!-\! 1\right)
   \left(\xi _1 \lambda _{xx}\!+\!2\right){}^2};\xi_0\right]. \label{eq:varphi1Dnontrivial}
\end{align}
Equation~\eqref{eq:varphi1Dnontrivial} displays advantageous properties  to complete the computation of number density  of extrema.  We have indeed
\begin{align}
\varphi_{\rm 1D}\left(\lambda _{\rho },\lambda _x,\lambda _{xx};\xi _0,\xi _1,\xi
   _2\right)&=\nonumber\\
   & 
   \hskip -2cm \varphi_{\rm 1D}\left(\lambda _{\rho }+\frac{\xi _1 \lambda _x^2}{2 \left(1-\xi _1 \lambda
   _{xx}\right)},0,\lambda _{xx};\xi _0,\xi _1,\xi _2\right)\,,
   \label{eq:D26}
\end{align}
which leads naturally to the change of variable
\begin{equation}
\lambda_{e}=\lambda _{\rho }+\frac{\xi _1 \lambda _x^2}{2 \left(1-\xi _1 \lambda
   _{xx}\right)}\,,
\end{equation}
when computing the inverse Laplace transformation over $\lambda_{x}$, necessary for the joint PDF, equation~\eqref{eq:formaldefPDF}.
We are left with
\begin{align}
&P(\rho,\rho_{x}=0,\rho_{,xx})=
\int\frac{\dd\lambda_{e}}{2\pi\ii}\frac{\dd\lambda _{xx}}{2\pi\ii}
\frac{(1-\xi_{1}\lambda _{xx})^{1/2}}{\sqrt{2\pi\xi_{1}\rho}}\nonumber\\
&\times
\exp\left(
-\lambda_{e}\rho-\lambda _{xx} \rho_{xx}\!+\!
\varphi_{\rm 1D}\left(\left\{\lambda _{e}, 0,\lambda _{xx}\right\},\xi _0,\xi _1,\xi
   _2\right)
\right)\,. \label{eq:D28}
\end{align}
We can further note that the dependence in $\lambda_{e}$ in $\varphi_{\rm 1D}\left(\left\{\lambda _{e}, 0,\lambda _{xx}\right\};\xi _0,\xi _1,\xi
   _2\right)$  now takes a very simple form
\begin{equation}
\varphi_{\rm 1D}\left(\left\{\lambda _{e}, 0,\lambda _{xx}\right\};\xi _0,\xi _1,\xi
   _2\right)=-\frac{2}{\xi _0}+\frac{a_{c}(\lambda _{xx})}{\lambda_{c}(\lambda _{xx})-\lambda_{e}}\,,
\end{equation}
with
\begin{align}
a_{c}(\lambda_{xx})&=\frac{\left(\xi _1 \lambda _{xx}+2\right){}^2}{\xi _0^2}\,,\\
\lambda_{c}(\lambda_{xx})&=\frac{1}{2\xi _0} \left(\xi _1 \lambda _{xx}+2\right){}^2-\frac{\xi _2}{2} \lambda
   _{xx}^2\,.
\end{align}
The integral over $\lambda_{e}$ in equation~\eqref{eq:D28} can then be done with the use of equation~\eqref{KeyLapInv}. It leads to the following expression 
(for the continuous, non singular, component)
\begin{align}
P(\rho,\rho_{x}=0,\rho_{xx})=
\int\frac{\dd\lambda _{xx}}{2\pi\ii}
e^{-\lambda_{xx}\rho_{,xx}}
\mixchi_{\mathrm{1D}}(\rho;\lambda_{xx})\,,
\end{align}
where
\begin{align}
&\mixchi_{\mathrm{1D}}(\rho;\lambda_{xx})=
\frac{2\sqrt{2}}{\xi_{0}^{2}\sqrt{\pi\xi_{1}\rho}}
{(1-\xi_{1}\lambda _{xx})^{1/2}}
{(1+\xi_{1}\lambda_{xx}/2)^{2}}\times 
 \label{mixchi1D}\\
&\hskip -0.15cm
\exp\left(\!-\frac{2}{\xi_{0}}\!+\!
\frac{\rho\xi_{2}}{2}\lambda_{xx}\!-\!
\frac{2\rho}{\xi_{0}}(1\!+\!\xi_{1}{\lambda_{xx}}/{2})^{2}\!\right)
\! \tHyp\!\!\left(2;\frac{4\rho}{\xi_{0}^{2}}  \,(1+\xi_{1}{\lambda_{xx}}/{2})^{2}\right).\notag
\end{align}
We are now in position to compute the number densities of extrema of the various kinds. The Euler
number density is obtained from the computation of equation~\eqref{1DnEuler}
which can be obtained directly via the derivative of $\chi_{\mathrm{1D}}(\rho,\lambda_{xx})$ defined in (\ref{mixchi1D}) with respect to $\lambda_{xx}$ at
position $\lambda_{xx}=0$,
\begin{equation}
n_{\rm 1D}^{\rm Euler}(\rho)=
\frac{\partial}{\partial\lambda_{xx}}_{\Big\vert_{\lambda_{xx}=0}}
\mixchi_{\mathrm{1D}}(\rho;\lambda_{xx})\,.
\end{equation}
This finally leads to the following remarkable closed form expression for the 1D Euler number density of Levy flights
\begin{align}
n_{\rm 1D}^{\rm Euler}(\rho)&=\sqrt{\frac{2\xi_1}{\pi\rho}}
\frac{e^{-\frac{2 (\rho +1)}{\xi _0}} }{
   \xi _0^3 }\nonumber\\
& \times  \left\{
2 \xi _0 \, \tHyp\left(1;\frac{4 \rho }{\xi
   _0^2}\right)-\left(\xi _0+4 \rho \right) \, \tHyp\left(2;\frac{4 \rho }{\xi _0^2}\right)
  \right\}. \label{eq:Euler1Dflight}
\end{align}
This is to be contrasted to the number density of Euler points in the Gaussian limit, which reads
\begin{equation}
n_{\rm 1D}^{\rm Gaussian}(\rho)=
\frac{\sqrt{\xi _1} (1-\rho) e^{-\frac{(\rho -1)^2}{2 \xi _0}}}{2\pi  \xi _0^{3/2}},
\end{equation}
which can also be obtained from more direct calculations \citep[see, e.g.][]{1986ApJ...304...15B}.  
In general,  it is of interest to be able to distinguish between the number counts of maxima and minima separately. This can be achieved via specific integration path in the complex plane as follows
\begin{equation}
n_{\rm 1D}^{\rm \pm}(\rho)=
\int_{-\ii\infty\pm\eps}^{+\ii\infty\pm\eps}\frac{\dd\lambda _{xx}}{2\pi\ii}
\frac{1}{\lambda_{xx}^{2}}
\mixchi_{\mathrm{1D}}(\rho;\lambda_{xx})\,,
\label{eq:P1Dextrema}
\end{equation}
where $\mixchi_{\mathrm{1D}}(\rho;\lambda_{xx})$ is given by equation~\eqref{mixchi1D}, while $\eps$ is a negative real constant for maxima and a positive real constant for minima\footnote{Note that the difference of the two extrema counts corresponds to a full integration around the origin which gives the back the Euler number density.}. This quadrature, together with its Euler counterpart, equation~\eqref{eq:Euler1Dflight}, is one of the key result of this paper.

\subsection{The 2D Euler and extrema counts}
\newcommand{\gamman}{{\gamma_n}}
\newcommand{\vgamma}{\vec{\gamma}}

In 2D, it is best to re-organise the field variables. Indeed
it is well known that while $\rho_{x}$ and $\rho_{y}$ form a vector field, $\kappa=(\rho_{xx}+\rho_{yy})/2$ is a scalar field and $\gamma_{1}=(\rho_{xx}-\rho_{yy})/2$ and $\gamma_{2}=\rho_{xy}$ form a spin-2 field.
It is then worth defining the CGF for those quantities and so to write it as a function of the conjugate variables 
of those components, that is respectively $\lambda_{\rho}$, $\lambda_{g_{1}}$, $\lambda_{g_{2}}$, $\lambda_{\kappa}$, $\lambda_{\gamma_{1}}$ and $\lambda_{\gamma_{2}}$. Furthermore, because the CGF must be rotation invariant, it can be expressed in terms of the following scalar quantities 
\begin{subequations}
    \begin{align}
\Lambda_{1}&=\lambda_{g_{1}}^{2}+\lambda_{g_{2}}^{2}\,,\\
\Lambda_{2}&=\lambda_{\kappa}=\lambda_{xx}+\lambda_{yy}\,,\\
\Lambda_{3}&=\lambda_{\gamma_{1}}^{2}+\lambda_{\gamma_{2}}^{2}=(\lambda_{xx}-\lambda_{yy})^{2}+\lambda_{xy}^{2}\,,\\
\Lambda_{4}&=
\lambda_{\gamma_{1}}(\lambda_{g_{1}}^{2}-\lambda_{g_{2}}^{2})+2\lambda_{\gamma_{2}}\lambda_{g_{1}}\lambda_{g_{2}}.
\end{align}
\end{subequations}
Following the same steps as in 1D, using {\sc \small Mathematica} to compute the small $d$ limit of the solution of the system of equations~\eqref{eq:tau9}, relying on the shifting relation (equation~\ref{scalecomposition}) and re-expressing the resulting generating function
in terms of the $\Lambda_i$s,
the final result for the generating function reads
\begin{align}
\varphi_{\rm 2D}&(\lambda_{\rho},\Lambda_{i};\xi_{0},\xi_{1},\xi_{2})=
\varphi_{1}
\left[\frac{\lambda _{\rho }+\frac{1}{6} \left(2 \Lambda _1^2+\Lambda _3\right) \xi _2}{\left(\frac{\Lambda _1 \xi
   _1}{2}+1\right){}^2}\right.
      \nonumber\\
&  -
\left.\frac{\xi _1 \left(\Lambda _2 \left(\Lambda _1 \xi _1-2\right)-\Lambda _4 \xi
   _1\right)}{\left(\frac{\Lambda _1 \xi _1}{2}+1\right){}^2 \left(\left(\Lambda _1^2-\Lambda _3\right) \xi _1^2-4
   \Lambda _1 \xi _1+4\right)};\xi_{0}\right] .   \label{eq:phi2Dcomplicated}
\end{align}
The calculation of the number density of extrema also follows the same steps as for
the 1D case. We first remark that
\begin{equation}
\varphi_{\rm 2D}\left(\lambda_{\rho},\Lambda_{i}\right)=
  \varphi_{\rm 2D}\left(\lambda _{e},\Lambda_{1}=0,\Lambda_{2},\Lambda_{3},\Lambda_{4}=0\right)\,,
\end{equation}
with
\begin{equation}
\lambda_{e}=\lambda _{\rho }+\frac{\xi_{1}(-2\Lambda_{1}+\xi_{1}\Lambda_{1}\Lambda_{2}-\xi_{1}\Lambda_{4})}{-4+4\xi_{1}\Lambda_{2}+\xi_{1}^{2}(\Lambda_{3}-\Lambda_{2}^{2})}\,,
\end{equation}
which allows us to integrate over $\lambda_{g_{1}}$ and $\lambda_{g_{2}}$. This leads to the joint PDF,
\begin{align}
P(\rho,\,&\rho_{i}=0,\rho_{ij})\!=\!
\int\frac{\dd\lambda_{e}}{2\pi\ii}\prod_{ij}\frac{\dd\lambda _{ij}}{2\pi\ii}
\frac{\left(
4-4\xi_{1}\Lambda_{2}+\xi_{1}^{2}(\Lambda_{2}^{2}-\Lambda_{3})
\right)^{1/2}}{4\pi\xi_{1}\rho}\nonumber\\
&\times
\exp\left(
-\lambda_{e}\rho- \sum_{ij}\lambda _{ij} \rho_{ij}+
\varphi_{\rm 2D}\left(\lambda _{e},0,\Lambda_{2},\Lambda_{3},0\right)
\right). \label{eq:PDFwithlambdae}
\end{align}
Then, defining
\begin{align}
a_{c}(\Lambda_{2})&=\frac{\left(\xi _1 \Lambda_2+2\right){}^2}{\xi _0^2}\,,\\
\lambda_{c}(\Lambda_{2},\Lambda_{3})&=\frac{1}{2\xi _0} 
\left(\xi _1 \Lambda_2+2\right)^2
-\frac{\xi _2}{6} (2\Lambda_2^2+\Lambda_{3})\,,
\end{align}
allows us to rewrite $\varphi_{\rm 2D}$ in equation~\eqref{eq:phi2Dcomplicated} as
\begin{equation}
\varphi_{\rm 2D}\left(\lambda _{e},0,\Lambda_{2},\Lambda_{3},0\right)=
-\frac{2}{\xi _0}+\frac{a_{c}(\Lambda_{2})}{\lambda_{c}(\Lambda_{2},\Lambda_{3})-\lambda_{e}}.
\end{equation}
Then we can carry out the integration over $\lambda_{e}$ in equation~\eqref{eq:PDFwithlambdae} with the help of (\ref{KeyLapInv}). This leads to the following expression for the joint PDF
\begin{eqnarray}
P(\rho,\rho_{i}=0,\rho_{,ij})=
\int \prod_{ij}\frac{\dd\lambda_{ij}}{2\pi\ii}
e^{-\lambda_{ij}\rho_{,ij}}
\mixchi_\mathrm{2D}(\rho;\Lambda_{2},\Lambda_{3})\,,
\end{eqnarray}
where
\begin{align}
\mixchi_\mathrm{2D}&(\rho;\Lambda_{2},\Lambda_{3})
=\frac{\left(\Lambda _2 \xi _1+2\right){}^2 \sqrt{\left(\Lambda _2^2-\Lambda _3\right) \xi _1^2-4 \Lambda _2 \xi _1+4}}
  {4 \pi    \xi _0^2 \xi _1 \rho }\nonumber\\
&\times   \exp \left(-\frac{2}{\xi_{0}}
   \left(\rho  \left\{\left(\frac{\Lambda _2 \xi _1}{2}+1\right){}^2-\frac{1}{12} \left(2 \Lambda _2^2+\Lambda _3\right) \xi _0 \xi
   _2\right\}+1\right)\right) \nonumber\\
&\times \ 
 \tHyp\left(2;\frac{\rho  \left(\Lambda _2 \xi _1+2\right){}^2}{\xi _0^2}\right)\,,
 \label{mixchi2D}
\end{align}
that has to be integrated over $\lambda_{\kappa}$, $\lambda_{\gamma_{1}}$ and $\lambda_{\gamma_{2}}$. 
In the following we note $\vgamma$ the two component vector $(\gamma_1,\gamma_2)$ 
and $\gamman$ its norm, $\gamma_n=\sqrt{\gamma_1^2+\gamma_2^2}$. We also note 
$\lambda_\gamma\equiv\sqrt{\lambda_{\gamma_1}^2+\lambda_{\gamma_2}^2}$.
The resulting number density of Euler points  follows through differentiation
\begin{align}
n&_{\rm 2D}^{\rm Euler}(\rho)\!=\!\int \dd\kappa\,\dd^{2}\vgamma \ (\kappa^{2}-\gamman^{2})\ 
P(\rho,\rho_{i}=0,\rho_{ij})\notag \\
&=
\left\{\left(\frac{\partial}{\partial\lambda_{\kappa}}\right)^{2}-
\left(\frac{\partial}{\partial\lambda_{\gamma_{1}}}\right)^{2}-
\left(\frac{\partial}{\partial\lambda_{\gamma_{2}}}\right)^{2}\right\}_{\Big\vert_{\lambda_{\rho}=\lambda_{\gamma_{1}}=\lambda_{\gamma_{2}}=0}}\!\!\!\!\!\!\!\!\!\!\!\!\!\!
\mixchi_\mathrm{2D}(\rho;\Lambda_{2},\Lambda_{3})\,,
\end{align}
which after some algebra yields the following final  closed form for the 2D Euler count of  \flights
\begin{align}
n_{\rm 2D}^{\rm Euler}(\rho)=&
-\frac{\xi _1 e^{-\frac{2 (\rho +1)}{\xi _0}} }{\pi  \xi _0^4 \rho }
\left\{\xi _0 \left(\xi _0+8 \rho \right) \,
   \tHyp\left(1;\frac{4 \rho }{\xi _0^2}\right)
   \right.\nonumber\\
&
   -\left.\left(2 \xi _0 \rho +\xi _0^2+8 \rho  (\rho
   +1)\right) \, \tHyp\left(2;\frac{4 \rho }{\xi _0^2}\right)\right\} 
   \label{eq:Euler2Dflight}\,,
\end{align}
to be once again contrasted to the Gaussian counts
\begin{align}
n_{\rm 2D}^{\rm Gaussian}(\rho)=
-\frac{\xi _1 e^{-\frac{(\rho -1)^2}{2 \xi _0}} \left(\xi _0-(\rho -1)^2\right)}{2 \sqrt{2} \pi ^{3/2} \xi_0^{5/2}}\,.
   \end{align}
The computation of extrema number densities in 2D require us to distinguish the signs of the eigenvalues of the Hessian. In practice that means one should be able to split the integral over $\kappa$ into 3 domains, $\kappa<-\gamman$, $-\gamman<\kappa<\gamman$ and $\kappa<\gamman$. 
We first note that
\begin{align}
\int_{-\gamman}^{\gamman}\!\!\!\dd\kappa(\kappa^2-\gamman^2)e^{\ii\kappa\lambda_\kappa}
\!=\!    -\frac{4}{\lambda_\kappa^3}\sqrt{\frac{\pi}{2}}\left(\gamman\lambda_\kappa\right)^{3/2}\!J_{3/2}\left(\gamma\lambda_\kappa\right),
\end{align}
which implies that 
\begin{align}
&\int_{\mathbb{R}^{2}}\dd^2\vgamma
\int_{-\gamman}^{\gamman}\dd\kappa(\kappa^2-\gamman^2)e^{\ii\kappa\lambda_\kappa+\ii\gamma_1 \lambda_{\gamma_1}+\ii\gamma_2 \lambda_{\gamma_2}}
\nonumber\\
&\hspace{1cm}  =-8\pi\sqrt{\frac{\pi}{2\lambda_\kappa^3}}
\int_0^\infty \dd\gamma\,\gamma^{5/2}
J_0\left(\gamma\lambda_\gamma\right)
J_{3/2}\left(\gamma\lambda_\kappa\right)\,,
\nonumber\\
&\hspace{1cm}  
=-12\pi(\lambda_\gamma^2-\lambda_\kappa^2)^{-5/2}\,,
\end{align}
which leads to the following quadrature for the extrema counts 
\begin{align}
    n_{\rm 2D}^{\rm extr.}(\rho)\!=\!
\int_{-\ii\infty+\eps}^{+\ii\infty+\eps}\!\!\frac{\dd\lambda_{\kappa}}{2\pi\ii}
\int_{0}^{\infty}\!\!\frac{\dd\lambda_{\gamma}}{2\pi}
\frac{12\pi\lambda_\gamma}{(\lambda_\gamma^2+\lambda_\kappa^2)^{5/2}}\ 
\mixchi_\mathrm{2D} (\rho;\Lambda_{2},\Lambda_{3})\,, \notag
\end{align}
where $\eps$ is a negative real constant for maxima and a positive real constant for minima. This double quadrature, together with its Euler counterpart \eqref{eq:Euler2Dflight}, is also one of the key result of this paper.

\subsection{The 3D Euler and extrema counts}

The 3D  Euler and extrema counts can be similarly derived, although the calculation is somewhat  more sophisticated. To sort out the results let us take full advantage of the rotation invariance of the result. Let us be a bit more precise.
The quantities, $\rho$, $\rho_{,i}$ and $\rho_{,ij}$ form respectively a tensor of rank 0, 1 and 2 with respect to coordinate changes $\mR_i^j$. Therefore we would like to derive the  conjugate expression 
$\hlambda_{i}$ and $\hlambda_{ij}$ of the quantities obtained after a change of coordinates,
    \begin{align}
    \hrho_{,i}&=\mR_i^{i'}\rho_{,i'}\,,\quad
    \hrho_{,ij}=\mR_i^{i'}\mR_j^{j'}\rho_{,i'j'}\,.
\end{align}
We should have 
\begin{align}
\sum_i\lambda_{i}\rho_{,i}=\sum_i\hlambda_{i}\hrho_{,i}\,,\label{scalarprod1}
\quad \sum_{i\ge j}\lambda_{ij}\rho_{,ij}=\sum_{i\ge j}\hlambda_{ij}\hrho_{,ij}.
\end{align}
Replacing $\hrho_{,i}$ by its expression in the right hand side of equation~\eqref{scalarprod1}  ensures that 
\begin{equation}
    \lambda_{i}=\mR_i^{i'}\hlambda_{,i'}\,,
\end{equation}
making clear that $\lambda_{i}$ is a rank-1 co-tensor. This is not the case for $\lambda_{ij}$ as the sum in equation~\eqref{scalarprod1} excludes repeated symmetric quantities. We are then led to define
\begin{equation}
    \lambdaR_{ij}=\left(
    \begin{array}{ccc}
    \lambda_{xx}&\lambda_{xy}/2&\lambda_{xz}/2\\
    \lambda_{xy}/2&\lambda_{yy}&\lambda_{yz}/2\\
    \lambda_{xz}/2&\lambda_{yz}/2&\lambda_{zz}
    \end{array}
    \right),\label{lambdaRdef}
\end{equation}
so that
\begin{equation}
\sum_{i\ge j}\lambda_{ij}\rho_{,ij}=\sum_{ij}\lambdaR_{ij}\rho_{,ij}.\label{scalarprod2R}
\end{equation}
It is then clear that $\lambdaR_{ij}$ transforms as a rank-2 co-tensor. From the definition of equation~\eqref{CGFdef1},
 the CGF itself is also clearly invariant under rotations. We then expect it to depend 
on combinations of $\lambda_{i}$ and $\lambdaR_{ij}$ that are scalar invariant.
Eventually we found the expression of the CGF to depend on the following six invariant quantities 
\begin{subequations}
    \begin{align}
\Lambda_{1}&=\sum_i \lambda_{i}\lambda_{i}\,,\quad
\Lambda_{2}=\sum_i \lambdaR_{ii}\,,\\
\Lambda_{3}&=\sum_{ij}\lambdaR_{ij}\lambdaR_{ij}\,,\quad
\Lambda_{4}=\det\left[\lambdaR_{ij}\right]\,,\\
\Lambda_{5}&=\sum_{ij}\lambdaR_{ij}\lambda_{i}\lambda_{j}\,,\,\,
\Lambda_{6}=\sum_{ijk}\lambdaR_{ij}\lambda_{i}\lambdaR_{jk}\lambda_{k}\,,
\end{align}
\end{subequations}
with 
\begin{align}
\varphi&_{\rm 3D}(\lambda_{\rho},\Lambda_{i};0,\xi_{1},\xi_{2})\!=\!
\left\{2 \xi _1^3 \left(12 \Lambda _4 \lambda _{\rho }\!+\!\Lambda _2^2 \left(2 \Lambda _4 \xi _2\!-\!3 \Lambda _1\right)\!+\!4 \Lambda _3 \Lambda _4 \xi _2
  \right)\right.\nonumber\\
& 
   \left. +2 \xi _1^3\left(6 \Lambda _5
   \Lambda _2+3 \Lambda _1 \Lambda _3-6 \Lambda _6\right)
   \right.
    \left. -2 \xi _1^2 \left(\Lambda _2^4 \xi _2-6 \Lambda _1 \Lambda _2+6 \Lambda _5\right)
   \right.
   \nonumber\\
&  
   \left. -2 \xi _1^2 \left(\Lambda _2^2 \left(6 \lambda _{\rho }+\Lambda _3 \xi _2\right)-2 \Lambda _3
   \left(3 \lambda _{\rho }+\Lambda _3 \xi _2\right)\right)
    \right.\nonumber\\
& 
\left.   +4 \xi _1 \left(6 \Lambda _2 \lambda
   _{\rho }\!+\!\Lambda _2^3 \xi _2\!+\!2 \Lambda _3 \Lambda _2 \xi _2\!-\!3 \Lambda _1\right)  \right.
\left. 
-4 \left(6 \lambda _{\rho }\!+\!\Lambda _2^2 \xi _2\!+\!2 \Lambda _3 \xi
   _2\right)
   \right\}\nonumber\\
&
\Large{/}\left\{6 \Lambda _2^2 \Lambda _4 \xi _1^5+3 \Lambda _2 \left(-\Lambda _2^3+\Lambda _3 \Lambda _2+8 \Lambda _4\right) \xi _1^4-24
   \right.\nonumber
   \\
&
\left. 
-\left(3 \left(2 \Lambda
   _2^3\!-\!4 \Lambda _2 \Lambda _3\right)\!-\!24 \Lambda _4\right) \xi _1^3\!+\!\left(6 \Lambda _2^2\!+\!12 \Lambda _3\right) \xi _1^2\right\},
\end{align}
and
\begin{equation}
\varphi_{\rm 3D}(\lambda_{\rho},\Lambda_{i};\xi_{0},\xi_{1},\xi_{2})=
\varphi_{1}(\varphi_{\rm 3D}(\lambda_{\rho},\Lambda_{i};0,\xi_{1},\xi_{2});\xi_{0}).
\end{equation}
As in 1 and 2D, the integral over $\lambda_i$ and $\lambda_\rho$ can be carried out explicitly.
The computation of the Euler number density 
from     equation~\eqref{eq:defeuler} relies on the following   
rules for the integration over $\lambda_{ij}$: 
$
    \det\left[\rho_{,ij}
    \right]\to
     \det\left[{\partial}/{\partial\lambda_{ij}}
    \right],
$
identifying $\lambda_{ij}$ and $\lambda_{ji}$.
After some significant algebra, we eventually get
\begin{align}
n_{\rm 3D}^{\rm Euler}(\rho)&=-\frac{e^{-\frac{2 (\rho +1)}{\xi _0}}\left(\xi _1 \rho \right){}^{3/2} }{2 \sqrt{2} \pi ^{3/2} \xi _0^5 \rho ^3}\nonumber\\
&\times
\left\{\left(6 \xi _0^2 \rho +16 \xi _0 \rho +3 \xi _0^3+32 \rho ^2 (\rho +3)\right)
   \, \tHyp\left(2;\frac{4 \rho }{\xi _0^2}\right)\right.\nonumber\\
&
   -\left.\left(16 \xi _0 \rho  (3 \rho +1)+3 \xi _0^3\right) \, \tHyp\left(1;\frac{4 \rho }{\xi
   _0^2}\right)\right\}\,,
\end{align}
while the corresponding Gaussian limit reads 
\begin{eqnarray}
n_{\rm 3D}^{\rm Gaussian}(\rho)=
-\frac{\xi _1^{3/2} (\rho -1) e^{-\frac{(\rho -1)^2}{2 \xi _0}} \left((\rho -1)^2-3 \xi _0\right)}{4 \pi ^2 \xi_0^{7/2}}\,.
\end{eqnarray}
This completes our results regarding Euler number densities.
Note that the 1,2 and 3D results are strikingly similar. Their Gaussian limit is consistent with the results presented 
in \citep{2013MNRAS.435..531C}, noting that $\xi_1=\langle |\nabla \rho|^2 \rangle/3$ and that 
the Euler characteristic and the genus differ by one integration.
\section{Extrema correlation}
\label{sec:extremaCorrelation}

The computation of the expected correlations of extrema is based on the derivation of the joint CGF for the density and its gradients at finite distance, 
\begin{equation}
    \varphi\left(\lambda_{\rho},\{\lambda_{i}\},\{\lambda_{ij}\};\mu_{\rho},\{\mu_{i}\},\{\mu_{ij}\}\right)\,.
\end{equation}
What makes its   estimation complicated is that it depends not only on the distance $d$, but also on the ratio of $d$ with the smoothing scale $R$. It is however possible to get a simple expression in the large separation limit. To carry out the calculation let us define $\xi_1(d)$ and $\xi_2(d)$ as respectively the first and second order derivative of the filtered densities at distance $d$, and Taylor expand w.r.t. $d$ 
\begin{equation}
    \xi(d+\delta l)\approx\xi_0(d)+\delta l\,\xi_1(d)+\frac{\delta l}{2}\xi_2(d).
\end{equation}
Using the same approach as before we can then derive joint CGF. 
An intermediate result is
given by the expression of the following quantity
\begin{align}
&\varphi^{c}(\lambda_{\rho},\lambda_{x},\lambda_{xx})\equiv
{\frac{\partial}{\partial\mu_\rho}}_{\Big\vert_{\mu_\rho=0}}\varphi(\lambda_{\rho},\lambda_{x},\lambda_{xx};\mu_{\rho})\nonumber\\
&=
\left\{-2 \xi_0(d) \lambda _{\rho}+\xi_1 \left(-\xi_0(d) \left(\lambda _x^2+\xi _2 \lambda
   _{\rm{xx}}^3-2 \lambda _{\rho } \lambda _{xx}\right)\right)\right.\nonumber\\
   &\hspace{.2cm}\left.
   +\xi_1 \left(\xi_2(d) \lambda _{xx}^2+\xi _0
   \left(\lambda _x^2+\xi_2 \lambda _{xx}^3-2 \lambda _{\rho } \lambda
   _{xx}\right)\right)\right.\nonumber\\
   &\hspace{.2cm}\left.
   -\xi_1(d) \lambda _x \left(\xi _1 \lambda _{xx}+2\right)
   +\xi _1^2
   \lambda _{xx}^2 \left(\xi_2(d) \lambda _{xx}+3\right)+\xi_0(d) \xi _2 \lambda
   _{xx}^2\right.\nonumber\\
   &\hspace{.2cm}\left.
   -2 \xi_2(d) \lambda _{xx}+2 \xi _0 \lambda _{\rho }+\xi _1^3 \lambda
   _{xx}^3-\xi _0 \xi _2 \lambda _{xx}^2-4\right\}^2\nonumber\\
&\hspace{.2cm}
\bigg{/}\left\{2 \xi _0 \lambda _{\rho }+\xi _0 \xi _1
   \left(\lambda _x^2+\xi _2 \lambda _{xx}^3-2 \lambda _{\rho } \lambda _{xx}\right)+\xi _1^3
   \lambda _{xx}^3+3 \xi _1^2 \lambda _{xx}^2\right.\nonumber\\
   &\hspace{.2cm}\left.-\xi _0 \xi _2 \lambda _{xx}^2-4\right\}^2\,,
\end{align}
which in principle can be used to build cross matter-extrema correlations at any separation\footnote{In practice however the inverse Laplace transformations have to be done numerically.}. 
It is worth investigating this expression in the large separation limit. We assume that $\xi_0(d)$, $\xi_1(d)$ and $\xi_2(d)$ are all small quantities. We can further note that 
$\xi_2(d)\xi_0\ll\xi_0(d)\xi_2$ if $d$ is much larger that the smoothing scale. We can then observe the following property in the large separation limit,
\begin{equation}
\varphi^c(\lambda_{\rho},\lambda_{x},\lambda_{xx})=1+
\varphi(\lambda_{\rho},\lambda_{x},\lambda_{xx})+
\xi_0(d)
\varphi(\lambda_{\rho},\lambda_{x},\lambda_{xx}).
\end{equation}
And more generally we can explicitly verify that
\begin{align}
\varphi(\lambda_{\rho},\lambda_{x},\lambda_{xx};\mu_\rho)&=
\varphi(\lambda_{\rho},\lambda_{x},\lambda_{xx})+
\varphi_1(\mu_\rho)\nonumber\\
&+
\varphi(\lambda_{\rho},\lambda_{x},\lambda_{xx})
\,\xi_0(d)\,\varphi_1(\mu_\rho)\,.
\end{align}
This is a property which is generic to MTM \citep{2022A&A...663A.124B}. More generally we expect that
equation~\eqref{eq:defvarphimultiple} holds.
Note that the correction  fully factorises in variables of each type.

In the large separation limit, the two-point number densities of critical points has correspondingly the following functional form
\begin{align}
  \hskip -0.1cm  n_{\rm crit.}(\rho_1;\rho_2)&\!=\!  n_{\rm crit.}(\rho_1) n_{\rm crit.}(\rho_1)
    \left\{
    1\!+\!b_{\rm crit.}(\rho_1)\xi_0(d)b_{\rm crit.}(\rho_2)
    \right\},
\end{align}
where $b_{\rm crit.}(\rho)$ is defined by 
\begin{align}
\hskip -0.75cm  b_{\rm crit.}(\rho)n_{\rm crit.}(\rho)\!=\!&
 \! \int\!\prod_{ij}{\dd\rho_{,ij}}
  {\rm Sgn}\left[\rho_{,ij}\right]
  \det\left[\rho_{,ij}\right]
    \! P_b\!\left(\rho,\rho_{,i}\!=\!0,\rho_{,ij}\right),\label{eq:bcritformal}
\end{align}
where 
${\rm Sgn}\left[\rho_{,ij}\right]$
is given in equation~\eqref{eq:defeuler}, and we have
\begin{align}
    P_b&\left(\rho,\rho_{,i}\!=\!0,\rho_{,ij}\right)\!=\!
    \!\! \int\!\frac{\dd\lambda_{\rho}}{2\pi\ii}
   \!\! \int\!\!\frac{\dd^{D}\lambda_{i}}{(2\pi\ii)^{D}}
    \!\!\int\!\!\prod_{ij}\frac{\dd\lambda_{ij}}{2\pi\ii}
    \varphi\left(\lambda_{\rho},\{\lambda_{i}\},\{\lambda_{ij}\}\right)\nonumber
\\
 &
    \times   
    \exp\left(-\ii\lambda_\rho\rho-\ii \sum_{ij}\lambda_{ij}\rho_{,ij}+
   \varphi\left(\lambda_{\rho},\{\lambda_{i}\},\{\lambda_{ij}\}\right)\right).
   \label{eq:bcritfull}
\end{align}
The calculation then follows the same articulation as before.

\subsection{1D bias asymptotic}
For the 1D case we are led to 
\begin{eqnarray}
P_b\left(\rho,\rho_{,x}=0,\rho_{,xx}\right)=
\int\frac{\dd\lambda _{xx}}{2\pi\ii}
e^{-\lambda_{xx}\rho_{,xx}}
\mixchi^\mathrm{1D}_b(\rho;\lambda_{xx})\,,
\end{eqnarray}
where
\begin{align}
\mixchi^\mathrm{1D}_b(\rho;\lambda_{xx})&=
\frac{2\sqrt{2}}{\xi_{0}^{3}\sqrt{\pi\xi_{1}\rho}}
{(1-\xi_{1}\lambda _{xx})^{1/2}}
{(1+\xi_{1}\lambda_{xx}/2)^{2}}
\nonumber\\
&\hspace{-1cm}\times
\exp\left(-\frac{2}{\xi_{0}}+
\frac{\xi_{2}}{2}\rho\,\lambda_{xx}-
\frac{2}{\xi_{0}}\rho\,(1+\xi_{1}\lambda_{xx}/2)^{2}\right)
\label{mixchib1D}\\
&\hspace{-1.7cm}\times
   \left\{\xi _0 \ \tHyp\left(1;\frac{4\rho}{\xi_{0}^{2}}  \,(1\!+\!\xi_{1}\lambda_{xx}/2)^{2}\right)
  \! -2 \ \tHyp\left(2;\frac{4\rho}{\xi_{0}^{2}}  \,(1\!+\!\xi_{1}\lambda_{xx}/2)^{2}\right)\right\}
\notag
\end{align}
that derives from the use of equations~(\ref{KeyLapInv}) and (\ref{KeyLapInv2}). It leads to the expression of the biased number density of Euler numbers
and of maxima. We thus have
\begin{align}
b&_{\rm Euler}^{\rm 1D}(\rho)n_{\rm Euler}^{\rm 1D}(\rho)
=
   \sqrt{\frac{2\xi_1}{\pi\rho}}
\frac{e^{-\frac{2 (\rho +1)}{\xi _0}} }{
   \xi _0^4 } \times \\
& \left\{
\xi _0\,(\xi_0 - 4 (1 + \rho)) \, \tHyp\left(1;\frac{4 \rho }{\xi
   _0^2}\right)
   +2 (\xi_0 + 8 \rho)  \, \tHyp\left(2;\frac{4 \rho }{\xi _0^2}\right)
  \right\}.\notag
\end{align}
which gives the behaviour of the large scale bias factor.
\subsection{2D bias asymptotic}
For the 2D case the corresponding form is 
\begin{align}
&\mixchi_b^\mathrm{2D}(\rho;\Lambda_{2},\Lambda_{3})= 
\frac{\left(\Lambda _2 \xi _1+2\right){}^2 \sqrt{\Lambda _2^2 \xi _1^2-\Lambda _3 \xi _1^2-4 \Lambda _2 \xi
   _1+4} }{4 \pi  \xi _0^3 \xi _1 \rho }
   \nonumber\\
&\times
   \exp \left(-\frac{3 \rho  \left(\Lambda _2 \xi _1+2\right){}^2-\left(2 \Lambda _2^2+\Lambda
   _3\right) \xi _0 \xi _2 \rho +12}{6 \xi _0}\right) 
    \label{mixchib2D}\\
&\times
   \left\{\xi _0 \ \tHyp\left(1;\frac{\rho 
   \left(\Lambda _2 \xi _1+2\right){}^2}{\xi _0^2}\right)\right.
\left.
   -2 \ \tHyp\left(2;\frac{\rho 
   \left(\Lambda _2 \xi _1+2\right){}^2}{\xi _0^2}\right)\right\}\,,
  \notag
\end{align}
which, after some algebra, leads  to  the bias factor
\begin{align}
\hskip -0.2cm b_{\rm Euler}^{\rm 2D}&(\rho)n_{\rm Euler}^{\rm 2D}(\rho)
\!=\!
\frac{2\xi_1 e^{-\frac{2 (\rho +1)}{\xi _0}} }{\pi  \xi _0^5 \rho }
\!\left\{\!\xi_0 (\xi_0 \!-\! 3 \xi_0 \rho \!+\! 4 \rho (3 \!+\! \rho)) \,
   \tHyp\left(1;\frac{4 \rho }{\xi _0^2}\right)
   \right.\nonumber\\
&
 \quad\quad  -\left.
   \left(\xi_0^2 + 8 \rho (1 + 3 \rho)\right) 
   \, \tHyp\left(2;\frac{4 \rho }{\xi _0^2}\right)\right\}.
\end{align}
The expressions (\ref{mixchi1D}), (\ref{mixchi2D}), (\ref{mixchib1D}) and (\ref{mixchib2D}) are the central building blocks for the construction of the extrema  correlations for the 1D and 2D cases.

\section{Gaussian CGFs} \label{sec:gauss}
The cumulant generating functions in their Gaussian limits the take the form
\begin{subequations}
    \begin{align}
\varphi&_{\rm 1D}^{\Gaussian}(\lambda_{\rho},\lambda_{x},\lambda_{xx})
=\lambda_{\rho}+\frac{1}{2}\xi_{0}\lambda_{\rho}^{2}+\frac{1}{2}\xi_{1}\lambda_{x}^{2}\nonumber\\
&\quad
-\xi_{1}
\lambda_{\rho}\lambda_{xx}+\frac{1}{2}\xi_{2}\lambda_{xx}^{2}\,,\\
\varphi&_{\rm 2D}^{\Gaussian}(\lambda_{\rho},\lambda_{i},\lambda_{ij})
=\lambda_{\rho}+\frac{1}{2}\xi_{0}\lambda_{\rho}^{2}+\frac{1}{2}\xi_{1}\sum_{i}\lambda_{i}^{2}\nonumber\\
&-\xi_{1}
\sum_{i}\lambda_{\rho}\lambda_{ii}
+\frac{1}{2}\xi_{2}\sum_{i}\lambda_{ii}^{2}
+\frac{1}{6}\xi_{2}\lambda_{xy}^{2}
+\frac{1}{3}\xi_{2}\lambda_{xx}\lambda_{yy}\,,\\
\varphi&_{\rm 3D}^{\Gaussian}(\lambda_{\rho},\lambda_{i},\lambda_{ij})
=\lambda_{\rho}+\frac{1}{2}\xi_{0}\lambda_{\rho}^{2}+\frac{1}{2}\xi_{1}\sum_{i}\lambda_{i}^{2}\nonumber\\
&-\xi_{1}
\sum_{i}\lambda_{\rho}\lambda_{ii}
+\frac{1}{2}\xi_{2}\sum_{i}\lambda_{ii}^{2}
+\frac{1}{6}\xi_{2}\left(\lambda_{xy}^{2}+\lambda_{yz}^{2}+\lambda_{xz}^{2}\right)
\nonumber\\
&+\frac{1}{3}\xi_{2}\left(\lambda_{xx}\lambda_{yy}+\lambda_{yy}\lambda_{zz}+\lambda_{xx}\lambda_{zz}\right)\,.
\end{align}
\end{subequations}

\label{lastpage}

\end{document}